\documentclass[sigconf,screen,nonacm]{acmart}

\AtBeginDocument{%
  }

\newcommand{\ProxyUntexSsim}{0.992}
\newcommand{\ProxyUntexFlip}{0.016}

\newcommand{\WarpFloatUntexSsimTwentyTwo}{0.948}
\newcommand{\WarpFloatUntexHolesTwentyTwo}{7.4}

\newcommand{\WarpThreeViewUntexSsimTwentyTwo}{0.977}
\newcommand{\WarpThreeViewUntexHolesTwentyTwo}{1.4}
\newcommand{\WarpFiveViewUntexSsimTwentyTwo}{0.982}
\newcommand{\WarpFiveViewUntexHolesTwentyTwo}{0.1}

\newcommand{\ProxyTexSsim}{0.964}
\newcommand{\ProxyTexFlip}{0.024}

\newcommand{\StaleTexSsimTwo}{0.797}
\newcommand{\WarpFloatTexSsimTwo}{0.961}

\newcommand{\StaleTexSsimTwentyTwo}{0.732}
\newcommand{\WarpFloatTexSsimTwentyTwo}{0.927}
\newcommand{\WarpFloatTexHolesTwentyTwo}{7.4}
\newcommand{\WarpTwelveFixTexSsimTwentyTwo}{0.796}

\newcommand{\WarpEightFixTexSsimTwentyTwo}{0.698}

\newcommand{\ProxyTermSsim}{0.957}
\newcommand{\ProxyTermFlip}{0.053}

\newcommand{\WarpTwelveFixTermSsimTwentyTwo}{0.718}

\newcommand{\WarpThreeViewTwelveTermSsimTwentyTwo}{0.777}

\newcommand{\WarpFiveViewTwelveTermSsimTwentyTwo}{0.751}

\newcommand{\MvBaselineMm}{450}

\newcommand{\DepthCodesEight}{2}

\newcommand{\DepthCodesTwelve}{11}

\newcommand{\HpsUntexSsim}{0.9910}

\newcommand{\HpsUntexPrims}{151}

\newcommand{\HpsUntexShOneDelta}{-0.0009}

\newcommand{\HpsUntexRefEightDelta}{-0.0056}

\newcommand{\HpsUntexRefThirtyTwoDelta}{-0.0023}

\newcommand{\HpsUntexRefResDelta}{-0.0007}

\newcommand{\HpsUntexInResDelta}{0.0007}

\newcommand{\HpsTexPrims}{117}

\newcommand{\HpsTexRefEightDelta}{-0.0111}

\newcommand{\HpsTexRefResDelta}{-0.0011}

\newcommand{\HpsTexInResDelta}{0.0238}
\newcommand{\HpsTexInResPrims}{362}

\newcommand{\HpsTermFlip}{0.0662}

\newcommand{\HpsTermShOneFlip}{0.0765}
\newcommand{\HpsTermShOneDelta}{-0.0053}

\newcommand{\HpsTermRefEightDelta}{-0.0066}

\newcommand{\HpsTermRefResDelta}{-0.0117}

\newcommand{\HpsTermInResDelta}{-0.0055}

\newcommand{\HpsTexOpSsim}{0.9643}

\newcommand{\HpsTexOpPrims}{362}

\newcommand{\WtToysOneKTwentyTwoSsim}{0.857}

\newcommand{\WtToysFiveKTwentyTwoSsim}{0.934}

\newcommand{\WtTermOneKTwentyTwoSsim}{0.747}

\newcommand{\WtTermOneKTwentyTwoHoles}{12.7}

\newcommand{\WtTermFiveKTwentyTwoSsim}{0.858}

\newcommand{\WtTermFiveKTwentyTwoHoles}{5.5}

\newcommand{\WtRotToysOneHoles}{31}
\newcommand{\WtRotToysOneSsim}{0.725}
\newcommand{\WtRotToysRotDeg}{16}

\newcommand{\WtRotToysFiveHoles}{20}
\newcommand{\WtRotToysFiveSsim}{0.863}

\newcommand{\WtDepthMbpsMin}{3.2}
\newcommand{\WtDepthMbpsMax}{14.4}

\newcommand{\DecUntexDepth}{0.002}
\newcommand{\DecUntexShading}{0.018}
\newcommand{\DecUntexDisocc}{0.048}
\newcommand{\DecUntexTotal}{0.068}
\newcommand{\DecUntexHoles}{7.7}
\newcommand{\DecUntexMm}{461}

\newcommand{\DecTexDepth}{0.009}
\newcommand{\DecTexShading}{0.044}
\newcommand{\DecTexDisocc}{0.032}
\newcommand{\DecTexTotal}{0.085}
\newcommand{\DecTexHoles}{7.7}
\newcommand{\DecTexMm}{461}
\newcommand{\DecTexClipDepth}{0.128}

\newcommand{\DecTermDepth}{0.003}
\newcommand{\DecTermShading}{0.067}
\newcommand{\DecTermDisocc}{0.104}
\newcommand{\DecTermTotal}{0.173}
\newcommand{\DecTermHoles}{8.2}
\newcommand{\DecTermMm}{230}
\newcommand{\DecTermClipDepth}{0.102}

\newcommand{\DecTermNearDepth}{0.003}
\newcommand{\DecTermNearShading}{0.081}
\newcommand{\DecTermNearDisocc}{0.131}
\newcommand{\DecTermNearTotal}{0.215}
\newcommand{\DecTermNearHoles}{11.5}
\newcommand{\DecTermNearMm}{230}

\usepackage{tikz}
\usetikzlibrary{positioning,fit,arrows.meta,calc}
\usepackage{graphicx}
\usepackage{subcaption}
\usepackage{comment}
\usepackage{enumitem}
\usepackage{xcolor}
\newcommand{\rev}[1]{#1}

\begin{document}

\title{Streaming Persistent 3D Gaussian Proxies for View-Flexible Remote Rendering}

\author{Matti Siekkinen}
\authornote{Both authors contributed equally to this research.}
\affiliation{%
  \institution{Aalto University}
  \city{Espoo}
  \country{Finland}
}
\email{matti.siekkinen@aalto.fi}

\author{Teemu Kämäräinen}
\authornotemark[1]
\affiliation{%
  \institution{University of Helsinki}
  \city{Helsinki}
  \country{Finland}}
\email{teemu.kamarainen@helsinki.fi}

\author{Ahmad Alhilal}
\affiliation{%
  \institution{Aalto University}
  \city{Espoo}
  \country{Finland}
}
\email{ahmad.alhilal@aalto.fi}

\renewcommand{\shortauthors}{Siekkinen, K\"am\"ar\"ainen, and Alhilal}

%\begin{abstract}
%   Cloud rendering is used in gaming and XR to overcome limitations in client side GPU computing capabilities and make it easier to use the applications with different devices. These systems mainly rely on streaming the rendered scene as 2D video. In this paper, we explore the use of 3D Gaussians as a means to convey real-time rendered scene graphics from server to clients. Our hypothesis is that 3D Gaussians offer better latency compensation through novel view synthesis than 2D video stream, and that continously optimizing a single 3D Gaussian model scales better in multi-user scenarios by avoiding the need to render, video encode, and stream each user’s view separately. We demonstrate the first results with a prototype system built for Unity.
%\end{abstract}

\begin{abstract}
Cloud rendering executes demanding real-time graphics on a remote server, but conventional video transmission constrains the displayed image to the server-rendered viewpoint. Image-space warping can compensate for client motion during transport and display latency, although its accuracy decreases with pose offset. We investigate an alternative approach in which the server constructs a persistent, client-renderable 3D Gaussian proxy online from game-engine render buffers and progressively transmits independently renderable spatial tiles. Our Unity prototype supports continuous optimization and expansion, approximate relighting, and engine-tracked rigid-object motion, while the client renders locally from its current viewpoint. We evaluate four scenes under controlled pose offsets and encoded walkthroughs, comparing the proxy with stale server-rendered frames, depth-assisted warping using HEVC with one, three, or five reference views. The proxy maintains stable quality as pose offset increases. At 50~Mbit/s, it achieves higher SSIM than all evaluated warping configurations at 367~ms on all four scenes and at 183~ms on three of the four scenes. The results identify complementary operating regimes: image warping is effective at small pose offsets, conventional video is rate-efficient when only the server-rendered viewpoint is required, while the persistent proxy provides robust viewpoint flexibility when the original renderer cannot be executed at the client.
\end{abstract}





\maketitle

\section{Introduction}
Interactive cloud rendering executes demanding graphics workloads on a remote server and typically delivers the rendered output as video~\cite{shi2015survey}. This enables thin and heterogeneous clients, but the displayed image remains tied to the viewpoint rendered by the server. When the client pose changes during render-to-display latency, image-space reprojection can update the view without another server render. Its accuracy, however, decreases as the viewpoint diverges from the transmitted image because disocclusions reveal missing content and view-dependent appearance no longer corresponds to the current viewpoint~\cite{mark1997postrendering}.

A client-renderable scene representation offers a different tradeoff. Once synchronized, the representation can be rendered directly from the current client pose rather than transporting pixels generated for an earlier viewpoint. 3D Gaussian Splatting (3DGS) is attractive for this role because its explicit primitives support real-time rasterization and can be progressively refined from image observations~\cite{kerbl20233dgs}. We refer to such a persistent, client-renderable approximation of the live scene as a \emph{3D Gaussian proxy}. The challenge is to construct the proxy while the application is running, keep it consistent as previously unseen regions and scene changes appear, and synchronize the evolving representation within a network budget.

This setting differs from conventional 3DGS streaming, where the representation is typically constructed beforehand or derived from captured content. Here, the running game engine provides privileged geometry, appearance, lighting, and object information from which the proxy can be initialized and continuously optimized. The resulting design occupies an intermediate point between two established remote rendering approaches. When only the server-rendered viewpoint is needed, video coding is highly rate-efficient. When the client can execute the application itself, asset delivery followed by state synchronization can reproduce the original renderer at low amortized transmission cost. A streamed proxy instead targets settings in which the original renderer or assets cannot be deployed at the client, but local rendering from viewpoints beyond the latest server pose is valuable.

We explore this design point with a Unity system that continuously constructs, expands, optimizes, and streams a persistent 3D Gaussian proxy. A native CUDA/Libtorch component initializes Gaussians from engine-native render buffers and refines them as the application runs. The evolving representation is partitioned into independently decodable spatial tiles with progressive refinements. Client pose and frustum feedback guide transmission, while received tiles remain cached in GPU memory and can be reused as the viewpoint changes. The representation also supports approximate relighting and engine-tracked rigid-object motion without retransmitting the complete scene state.

We evaluate the system in four scenes using both controlled pose-offset experiments and rate-limited streaming walkthroughs. We compare the proxy with stale server frames, depth-assisted reprojection using one, three, or five reference views, and low-latency video coding. The experiments reveal distinct operating regimes. Image-based reprojection provides high fidelity at small pose offsets but degrades as displacement exposes missing geometry and stale view-dependent appearance. In contrast, the client-rendered proxy maintains approximately stable quality as pose offset increases. At a 50 Mbit/s rate ceiling, the proxy has higher measured SSIM than all evaluated warping configurations at 367 ms on all four scenes, and at 183 ms on three of the four scenes. These results position a persistent 3D Gaussian proxy as a view-flexible remote-rendering layer between viewpoint-specific video delivery and full client-side application execution.

The main contributions of this work are:
\begin{itemize}[leftmargin=*]
    \item To our knowledge, the first online game-engine-to-3DGS pipeline that constructs, continuously expands and optimizes a persistent client-renderable scene proxy from engine-native observations, with support for approximate relighting and engine-tracked rigid-object motion.
    \item a rate-controlled streaming architecture for an evolving 3DGS representation, using independently decodable spatial tiles, progressive refinements, client-view-driven selection, and persistent client-side caching.
    \item a systematic evaluation across four scenes that characterizes convergence, pose-offset robustness, and rate--quality tradeoffs against depth-assisted warping using low-latency video.
\end{itemize}

\section{Background and Related Work}

Remote rendering traditionally sends server-rendered images or video to thin clients~\cite{shi2015survey}. Post-rendering reprojection can compensate for small pose changes without another server render, but it cannot recover disoccluded geometry or update view-dependent appearance~\cite{mark1997postrendering,bao2003lowbandwidth,
evangelakos2016timewarp,kim2020latewarp,gul2022latencywarping}. Hybrid systems instead split work between server and client, for example by combining cloud-computed lighting with local rasterization~\cite{crassin2015cloudlight,tan2024dhr+}. These approaches reduce latency sensitivity but still depend on a client-side renderer or viewpoint-specific image data.

A second line of work transmits lightweight scene proxies that support local view synthesis. Camera Offset Space and Trim Regions compute online visibility regions for streaming rendering~\cite{hladky2019cos,voglreiter2023trim}. QuadStream transmits layered, view-aligned quads derived from G-buffers~\cite{hladky2022quadstream}, while QUASAR adds a compact representation, temporal compression, and bandwidth adaptation for thin clients~\cite{lu2025quasar}. QuadStream and QUASAR are particularly close to our remote-rendering objective, but their transmitted representations are constructed around view-dependent or per-frame scene information. Our system instead maintains a persistent 3D representation that is refined and reused across frames and viewpoints.

3DGS represents a scene with anisotropic primitives optimized from multi-view observations and rendered in real time
~\cite{kerbl20233dgs}. Subsequent work has addressed persistent dynamic Gaussians~\cite{luiten2024dynamic3dgs}, online reconstruction and free-viewpoint streaming~\cite{sun20243dgstream,yan2025igs,huang2025echoes}, and Gaussian-based
immersive-media streaming for XR~\cite{shi20253dgsstreamingxr}. Compression and streaming methods exploit quantization, temporal redundancy, spatial partitioning, or scalable representations for dynamic Gaussian content
~\cite{girish2024queen,hu20254dgc,yin2026cags,naqvi2026cubegs,wang2024v,gong2025adaptive}. Other work reduces representation size or organizes Gaussians for efficient playback ~\cite{papantonakis2024reduced3dgs,niedermayr2024compressed3dgs, kerbl2024hierarchical3dgs} .%Recent thin-client Gaussian streaming also explores adaptive delivery over HTTP/3~\cite{artioli2026thin}.
 These systems primarily target captured, reconstructed, or pre-existing Gaussian content. In contrast, we construct and continuously maintain the representation from privileged geometry, appearance, lighting, and object information produced by a running game engine, while the application itself remains interactive.

Relightable Gaussian methods augment the representation with normals, material parameters, illumination, or visibility to support editable appearance and shadows~\cite{gao2024relightable3dgs,jiang2024gaussianshader,zhenyuan2025bigs,bai2025gare}. Our objective is narrower: we use an engine-compatible diffuse lighting model and approximate visibility that remain practical during continuous online optimization. Dynamic 3DGS methods address more general time-varying geometry, whereas our current system exploits game-engine object identity and explicit rigid transforms to handle motion.

When the client can execute the original application, compressed asset delivery followed by state synchronization provides a natural native-execution reference point, provided that intellectual property restrictions do not prohibit it. Our method targets the intermediate setting between native application delivery and viewpoint-specific image or per-frame proxy delivery: it constructs a persistent Gaussian representation from a live Unity application, progressively synchronizes that representation under a rate constraint, and renders it locally from the current client pose.

\section{System Design}
%Figure~\ref{fig:pipeline} should provide an overview of the system. At a high level, the server runs a Unity application together with a native CUDA/C++ plugin. A set of reference cameras observe the scene around the current player or viewer position. For each selected camera, Unity renders RGB images and additional buffers that expose geometric and shading information, such as world-space position, normal, depth, and spherical-harmonic (SH) transfer coefficients. These buffers are copied to native GPU resources and mapped into tensors in the plugin.

%The native plugin maintains the current Gaussian model and performs the learning step. It uses the incoming reference observations to initialize new Gaussians, update attributes of previously created Gaussians, and render differentiable predictions for optimization. At the same time, the same model can be rendered for inference or exported to a client-side viewer. The system therefore alternates between optimization and rendering while preserving a single evolving scene representation.

%\begin{figure}[t]
%    \centering
%    \fbox{\parbox{0.95\linewidth}{
%    Placeholder for pipeline figure: Unity scene $\rightarrow$ reference-camera rendering $\rightarrow$ RGB/G-buffer/SH extraction $\rightarrow$ native CUDA plugin $\rightarrow$ Gaussian optimization $\rightarrow$ client/viewer rendering.
%    }}
%    \caption{High-level system overview.}
%    \label{fig:pipeline}
%\end{figure}

Our system couples the game engine with a continuously optimized 3D Gaussian representation. We describe its Unity implementation, online model construction and expansion, support for relighting and rigid motion, and progressive streaming to a remote client.

\subsection{Overview}

% \rev{Figure~\ref{fig:system_overview} shows the end-to-end system architecture. Unity simulates the scene, manages cameras, and produces input textures. A native C++/CUDA plugin stores and optimizes the Gaussian tensors, renders the current model, and serializes it for transmission.}

% Preamble:
% \usepackage{tikz}
% \usetikzlibrary{positioning,fit,arrows.meta,calc}

%Overview of the live engine-driven 3DGS pipeline. On the server, Unity provides engine-native scene observations to a native plugin that initializes, optimizes, renders, and streams a 3D Gaussian Splatting representation. The client reconstructs the streamed model from the model snapshots, tensor deltas, and transform updates, renders local viewpoints from the received representation, and sends its current camera pose back to the server.

\begin{figure}[t]
    \centering
    \includegraphics[width=\linewidth]{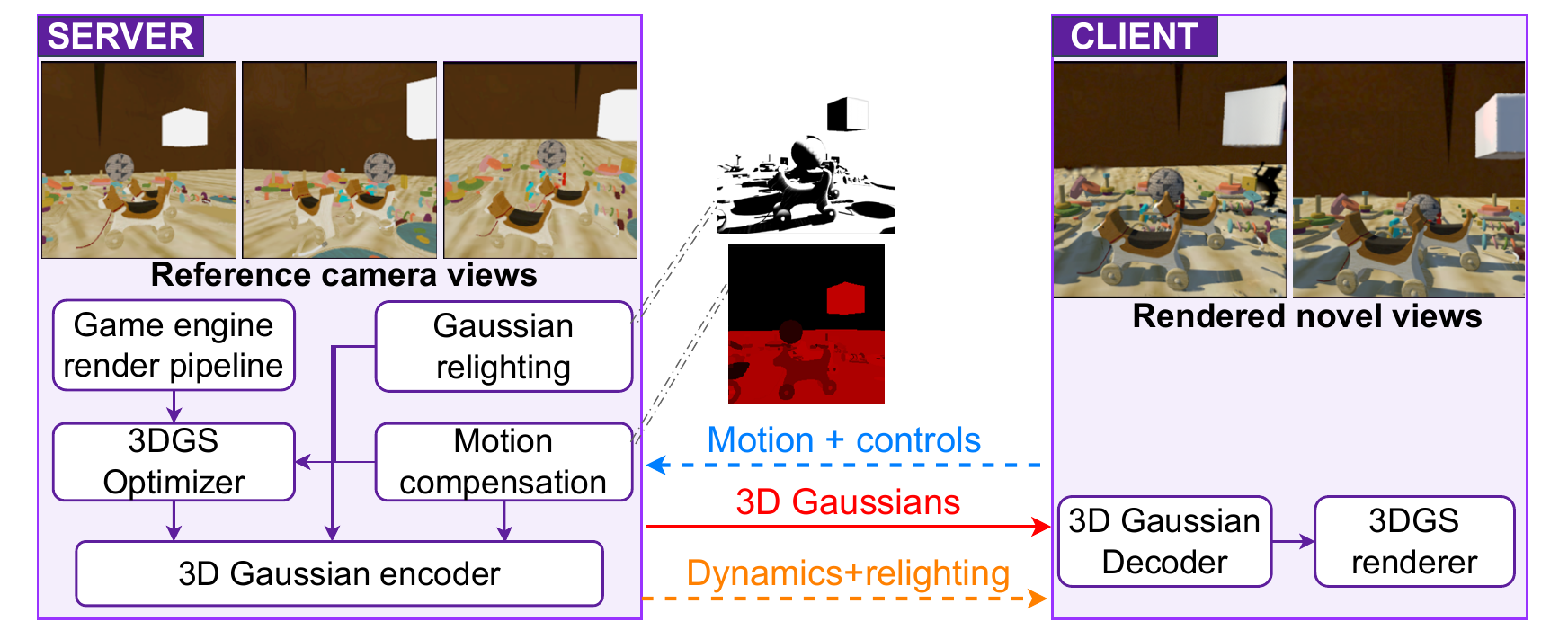}
    \caption{System overview. The server constructs and transmits spatially tiled Gaussian data; the client caches the tiles, renders its current viewpoint, and returns camera and frustum state.}
    \label{fig:system_overview}
\end{figure}

Figure~\ref{fig:system_overview} illustrates the end-to-end system. On the server, Unity runs the application and provides engine-native geometry, appearance, lighting, and object information to a native C++/CUDA plugin. The plugin maintains the evolving 3D Gaussian proxy, initializes and optimizes its primitives, and prepares the representation
for transmission.

The client receives and caches the streamed Gaussian data and renders the representation from its current viewpoint. Camera pose and frustum feedback are returned to the server and guide both scene acquisition and transmission. Optimization runs concurrently with the Unity render loop, allowing new observations, object motion, and lighting updates to be incorporated while the representation remains available for streaming and rendering.

\subsection{Online Gaussian Model Construction and Optimization}
\label{sec:construction}

Model construction has three stages. Unity first acquires engine-native observations of the scene. These observations initialize new Gaussian primitives, which are then refined through differentiable rendering. Low-resolution \emph{input cameras} provide dense samples for initialization, while a smaller set of higher-resolution \emph{reference cameras} provides supervision for optimization.

\subsubsection{Input-View Acquisition}
The input cameras form a dome around the active region, centered at the tracked client position and oriented with its heading. This limits the
number of views while providing coverage of the viewer's current
neighborhood.

%For each input camera, the server renders not only the geometric information needed to initialize new 3D Gaussians but also several auxiliary buffers required for appearance, relighting, and scene dynamics. A dedicated world-space point pass stores per-pixel 3D positions, and depth-based neighborhood differences are used to estimate an initial world-space scale for the corresponding Gaussian primitive. Additional passes provide surface normals, lighting-related quantities used to initialize spherical-harmonics (SH) appearance coefficients, and object IDs for associating rigid scene objects with the resulting 3D Gaussians. This data is written to GPU-accessible buffers and copied to native resources that are subsequently imported into CUDA memory for further processing, keeping the initialization pipeline entirely in GPU memory. Together, these buffers provide the engine-native input from which the initial Gaussian representation is formed.

Each input view provides world-space positions, surface normals, appearance and lighting quantities, and object IDs. Local depth differences are used to estimate the initial spatial footprint of each sample. Where input views overlap, the system retains the least-oblique observation. The selected samples remain in GPU-accessible buffers and are passed directly to CUDA for Gaussian initialization.

\subsubsection{Gaussian Initialization from Engine-Native Inputs}
The plugin maps sampled world-space positions to Gaussian means and the estimated footprints to isotropic log scales. Appearance is initialized as spherical-harmonic coefficients, and each new primitive is associated with the corresponding object ID and transform when applicable. The resulting batch is appended to the active Gaussian pool to immediately participate in optimization and streaming.

Because the geometry and appearance observations come directly from the running game engine, initialization does not require an external structure-from-motion or photogrammetry stage. This also allows newly observed regions to be populated while the application is running.

\subsubsection{Continuous Optimization}
The plugin renders the current Gaussian model from the reference cameras and compares these predictions with the corresponding Unity reference images. The resulting image-space loss updates the standard 3DGS parameters through differentiable rendering. During initial construction, density control duplicates, splits, and prunes primitives and periodically resets opacity.

Newly observed regions are not populated through density control alone. Instead, engine-derived samples directly initialize new Gaussians, which then enter the same optimization process as the existing representation. As the viewer explores the scene, reconstruction quality therefore depends on both the accumulated optimization time and the rate at which new regions become visible.

\subsection{Progressive Model Expansion}

A regular 3D grid partitions the scene and provides the unit of progressive expansion. As the client moves, the server considers cells around the active region and tests their visibility from the input cameras. Visible cells that are not yet sufficiently represented are initialized using the same engine-derived procedure described in Section~\ref{sec:construction}. The resulting Gaussians enter the active pool and immediately participate in optimization.

To prevent optimization cost from growing with the explored scene, \emph{Gaussian freezing} separates actively optimized content from settled regions. Active Gaussians retain gradients and optimizer state. Once a region is considered settled, its Gaussians move to a frozen, inference-only pool. Frozen Gaussians remain part of the renderable scene but no longer require optimizer state or gradient updates.

\emph{Preculling} further limits the work performed for each view. Camera-based depth tests identify the grid cells relevant to the current view, and only Gaussians from those cells enter rendering and optimization. Together, progressive insertion, freezing, and preculling allow the representation to expand with scene exploration without requiring the full accumulated model to remain active in every optimization step.

%This cell-based formulation is useful for the paper because it connects directly to scalability. It provides a natural unit for reporting scene coverage, optimization progress, and model growth, and it can later be reused by the networking layer for spatial prioritization and transmission scheduling.

%\subsection{Scene Dynamics}
\subsection{Relighting and Rigid-Object Motion}
\label{sec:relight_dyn}
%\subsubsection{3D Gaussian Relighting}
%\rev{Each Gaussian stores diffuse SH coefficients that are optimized online and combined with Unity's current SH lighting at render time. This low-frequency transfer representation responds to light direction and intensity without reconstructing the model.}

%\rev{For directional shadows, an orthographic light camera produces a depth map. Projecting Gaussian centers into this map yields a per-Gaussian visibility value that attenuates direct light. The method captures coarse shadow changes, not full global illumination.}

\subsubsection{3D Gaussian Relighting}
Each Gaussian stores diffuse spherical-harmonic (SH) transfer coefficients that are optimized together with the representation. At render time, these coefficients are combined with Unity's current SH lighting, allowing the rendered appearance to respond to changes in light direction and intensity without reconstructing the Gaussian model.

Directional shadows are approximated using a light-space depth map rendered from an orthographic camera. Gaussian centers are projected into this map to estimate a per-Gaussian visibility term, which attenuates direct illumination. This captures coarse shadow changes, but does not model higher-frequency visibility or global illumination.

%\subsubsection{Dynamic Object Handling}
%\rev{Gaussians attached to a rigid object retain its ID and object-local coordinates. Before rendering or optimization, Unity's tracked transform maps them into world space, so image-space loss need not recover object motion.}

%This mechanism is particularly important for objects whose motion is not well modeled as appearance change alone. By updating transforms explicitly, the system preserves a stable mapping between Gaussians and the physical entities they represent. The same infrastructure also supports recording and replay of object motion together with lighting changes, which is helpful for controlled experiments.

\subsubsection{Dynamic Object Handling}
While initializing Gaussians, Unity assigns persistent IDs to tracked rigid objects and renders these IDs alongside the other per-pixel Gaussian initialization attributes. Each Gaussian therefore inherits the ID of the object from which it was sampled. The plugin also maintains object-local means and orientations for these Gaussians, while Unity provides the current rigid transform for each tracked object.

Before rendering or optimization, the corresponding object transform maps the local Gaussian state into world space. For optimization, the object pose captured with each reference image is applied to keep the Gaussian model registered with the moving object. After geometry updates, the object-local Gaussian parameters are refreshed. The same object IDs and transforms are used by the streaming protocol (Section \ref{sec:streaming_design}), allowing rigid motion to be updated without retransmitting the Gaussian attributes.

%\subsection{Streaming the Live 3DGS Representation}
\subsection{Streaming the Evolving 3DGS Representation}
\label{sec:streaming_design}

\begin{figure}[t]
    \centering
    \definecolor{srpurple}{HTML}{7030A0}
    \definecolor{srred}{HTML}{D7191C}
    \definecolor{srblue}{HTML}{0070C0}
    \definecolor{srorange}{HTML}{F07818}
    \begin{tikzpicture}[
        font=\scriptsize,
        >=Latex,
        panel/.style={draw=srpurple, fill=srpurple!6, rounded corners=2pt,
            line width=0.8pt},
        stage/.style={draw=srpurple!85!black, fill=white, rounded corners=1.5pt,
            align=center, inner sep=2.2pt, minimum height=13mm, text width=18.5mm},
        heading/.style={fill=srpurple, text=white, rounded corners=1.2pt,
            font=\scriptsize\bfseries, inner xsep=4pt, inner ysep=2pt},
        note/.style={draw, fill=white, rounded corners=1.2pt, align=center,
            inner sep=2pt, text width=20mm},
        internal/.style={->, draw=srpurple!80!black, line width=0.8pt},
        tiles/.style={->, draw=srred, line width=1.0pt},
        feedback/.style={->, draw=srblue, dashed, line width=0.9pt},
        motion/.style={->, draw=srorange, densely dashed, line width=0.9pt}
    ]
        % Explicit server and client regions keep the two sides visually distinct.
        \node[panel, minimum width=8.15cm, minimum height=2.15cm] (serverpanel) at (3.35,0) {};
        \node[heading, anchor=west] at (-0.57,0.82) {SERVER};
        \node[stage] (snapshot) at (0.45,0) {\textbf{Model snapshot}\\[-1pt]
            {\tiny Gaussian parameters}};
        \node[stage, text width=20mm] (coding) at (3.15,0) {\textbf{Tile encoder}\\[-1pt]
            {\tiny base + refinements\\adaptive coding + Zstd}};
        \node[stage, text width=20mm] (schedule) at (6.05,0) {\textbf{View and rate\\scheduler}\\[-1pt]
            {\tiny visible bases first}};
        \draw[internal] (snapshot) -- (coding);
        \draw[internal] (coding) -- (schedule);

        \node[panel, minimum width=8.15cm, minimum height=1.85cm] (clientpanel) at (3.35,-4.05) {};
        \node[heading, anchor=west] at (-0.57,-3.38) {CLIENT};
        \node[stage, text width=22mm] (cache) at (1.65,-4.05)
            {\textbf{Receiver and\\tile cache}};
        \node[stage, text width=22mm] (render) at (5.05,-4.05)
            {\textbf{Local Gaussian\\renderer}};
        \draw[internal] (cache) -- (render);

        % The centre lane separates data, motion, and feedback traffic.
        \node[note, draw=srorange, text width=19mm] (objects) at (0.45,-2.1)
            {\textbf{Rigid-object updates}\\[-1pt]{\tiny object ID + transform}};
        \node[note, draw=srred, text width=21mm] (tilelabel) at (3.15,-2.1)
            {\textbf{Gaussian tiles}\\[-1pt]{\tiny rate-controlled}};
        \node[note, draw=srblue, text width=20mm] (viewfeedback) at (6.05,-2.1)
            {\textbf{Client feedback}\\[-1pt]{\tiny camera + frustum\\target rate + limits}};

        \draw[motion] (snapshot.south) -- (objects.north);
        \draw[tiles] (schedule.south) -- (tilelabel.north);
        \draw[tiles] (tilelabel.south) -- (cache.north);
        \draw[motion] (objects.south) -- ($(cache.north)+(-0.65,0)$);
        \draw[feedback] (render.north) -- (viewfeedback.south);
        \draw[feedback] (viewfeedback.north) -- (schedule.south);
    \end{tikzpicture}
    \caption{Streaming pipeline. Server snapshots are encoded as independently renderable, versioned spatial tiles. Client feedback controls tile selection and rate, while received tiles are cached for reuse. Rigid-object transforms are sent separately so that motion does not require recoding Gaussian attributes.}
    \label{fig:streaming_design}
\end{figure}

Figure~\ref{fig:streaming_design} shows the streaming pipeline. An asynchronous worker reads coherent snapshots of the optimized Gaussian tensors so that encoding and transmission do not block the optimizer. Gaussians are partitioned into spatial tiles in object-local coordinates. Each tile is identified by its object, spatial cell, and bounded shard, which keeps tile identities stable even when tensor row ordering changes.

Each tile also carries an epoch and content version so that updates can be announced, replaced, or removed independently. Changes to topology or Gaussian attributes therefore affect only the corresponding tiles rather than requiring a new whole-scene snapshot.

Each tile begins with an independently renderable base layer so that the client can use it immediately after reception. The base quantizes position, log scale, orientation, opacity, and degree-zero SH appearance. Optional geometry and appearance refinements encode residuals relative to the base, with the appearance refinement also carrying the configured higher-order SH coefficients.

For each residual block, the encoder selects among dense, bitset-sparse, and delta-indexed layouts and applies Zstd only when it further reduces the encoded size. This extends the progressive tiling principle of CAGS~\cite{yin2026cags} to snapshots produced continuously by an online optimizer.

At session setup, the client and server negotiate the codec configuration, SH degree, message-size limits, and cache bounds. During streaming, the client reports its current camera pose, object-local view frusta, target rate, burst allowance, and visible/prefetch limits. The reported pose can also steer the client-conditioned acquisition cameras on the server.

The scheduler prioritizes missing visible base tiles, followed by visible replacements, nearby prefetch bases, and refinements. Newer unsent updates supersede older ones. Once transmission of a base or refinement begins, however, its remaining chunks are completed before another transaction starts. A shared byte-token bucket limits all downstream application-protocol traffic, including tile data and rigid-motion control messages.

The client validates tile framing, versions, checksums, and complete transactions before committing received data. Decoded bases and refinements are stored in a bounded cache, allowing previously received but currently invisible tiles to be reactivated without retransmission. The client adopts newly committed tile updates at most once every
100~ms by default, while rendering continues every frame from the latest local camera pose using the current assembled model.

Rigid-object motion uses a separate latest-wins control path. Gaussian tiles remain in canonical object-local coordinates, while object IDs associate each tile with a streamed absolute rigid transform. The client applies the newest transform to all cached tiles belonging to that object. Rigid translation or rotation therefore does not require retransmitting the Gaussian attributes.

Changes to geometry, topology, or appearance still require tile replacement or refinement. Deformable motion is not supported by the current protocol.

%\rev{Scene acquisition combines fixed coverage with views conditioned on the client. Sixteen authored cameras remain fixed, while four additional cameras are assigned to the active client. During a 30-s initialization phase, these cameras orbit the region in front of the initial client pose to establish multi-view coverage. When interaction begins, one camera follows the client pose exactly and three maintain nearby offsets. The same client feedback drives both acquisition and view-dependent tile selection, and benchmark camera input is disabled to ensure reproducibility.}

\section{Experimental Setup}
\label{sec:evaluation}

The evaluation has two parts. Section~\ref{sec:quality} examines view synthesis under controlled pose offsets, continuous optimization, progressive expansion, and scene dynamics. Section~\ref{sec:streaming} evaluates the complete rate-constrained streaming system against depth-assisted warping.

\subsection{System Setup}

\textbf{Scenes.} We evaluate four scenes. Toys and Textured toys share geometry and lighting but differ in surface detail. The textured variant provides a more discriminative reconstruction test. Terminal is taken from Unity's URP sample project. Terminal foreground adds 10-cm objects at a minimum distance of 0.45~m from the camera path to increase near-field disocclusion. Dynamic-scene experiments additionally use a bouncing textured ball, a rotating cube, and a rotating directional light.

\textbf{Trajectories.} The controlled experiments of Section~\ref{sec:quality} use a 30-s constant-speed orbit, allowing delay to be mapped directly to pose offset. The streaming experiments of Section~\ref{sec:streaming} use 60-s closed-spline walkthroughs with varying translation speed and heading. Toys and Textured toys share one walkthrough, while Terminal and Terminal foreground share another.
%\rev{The Toys walkthrough reuses the trajectory from the original streaming experiment. Both Terminal variants use the same walkthrough to isolate the effect of scene content. The controlled-orbit experiments on the toy scenes use a fixed directional light, whereas the streaming experiments retain each scene's authored lighting.}

\textbf{Camera rigs.} Input cameras form a Fibonacci dome around the active region. The reference-camera rig is a similar dome arrangement around the viewer (the client player camera pose) and follows it at all times, which keeps the optimization views aligned with the region currently being explored. Individual experiments specify their camera counts and resolutions.

\textbf{Metrics.} To assess visual quality, we use mean structural similarity (SSIM) and mean FLIP~\cite{andersson2020flip} relative to Unity ground truth rendered from the same viewpoint under the same lighting. Additional experiment-specific metrics are introduced where used.

\textbf{Hardware.} View-synthesis experiments use an NVIDIA GeForce RTX~4070 SUPER with 12~GB of VRAM. Streaming experiments use an RTX~3080~Ti server and an RTX~A2000 client connected over a local network.

\subsection{Baselines} 
\label{sec:baselines}
We compare the 3DGS proxy to several video-based baselines. The client pose is $P_c=\mathrm{pose}(t)$ and the most recent server pose is $P_s=\mathrm{pose}(t-k/60)$ for $k\in\{0,2,4,11,22\}$ frames, corresponding to $\{0,33,67,183,367\}$\,ms. 
%On Toys, these delays produce displacements of $\{0,\OffsetMmUntexTwo,\ldots,\OffsetMmUntexTwentyTwo\}$\,mm. 
%All baselines are evaluated against ground truth at $P_c$ with identical display encoding and $4\times$ multisampling.

\textbf{Stale frame}: the image rendered at $P_s$, displayed unchanged

\textbf{Depth-assisted warp:} the server image forward-warped to $P_c$ as a z-buffered point splat. Single-pixel cracks are filled and disocclusions remain empty. Depth is lossless or inverse-depth quantized to 12 or 8 bits over 0.01--100\,m

\textbf{Multi-view plus depth:} the server view plus views translated \MvBaselineMm\,mm left/right (three views) and up/down (five views). Simultaneous, equally oriented views are merged into one client z-buffer, using lossless or 12-bit depth

\section{View-Synthesis Evaluation}
\label{sec:quality}

\subsection{Parameter Sensitivity}
\label{sec:eval_params}

We first select a baseline configuration for each scene through hyperparameter search because the converged model quality is highly dependent on the parameterization. We vary the SH degree (1--3), number of reference views (8, 16, or 32), reference resolution ($1280^2$ or $1808^2$), initial Gaussian coverage (1.0--2.5~px), and input resolution ($256^2$ or $512^2$), considering also the constraint set by available GPU memory. The search sweeps through the parameter values and compares the resulting model quality with selected novel views.

\textbf{Reference-view count:} Among the tested values, sixteen views provide the highest SSIM on Toys: using eight or thirty-two views changes SSIM by $\HpsUntexRefEightDelta$ and $\HpsUntexRefThirtyTwoDelta$, respectively. Eight views change SSIM by $\HpsTexRefEightDelta$ on Textured toys and $\HpsTermRefEightDelta$ on Terminal. A small view set provides insufficient constraints, whereas a larger set reduces the optimization frequency per view under a fixed processing budget.

\textbf{SH degree:} Reducing the SH degree from three to one changes SSIM by $\HpsUntexShOneDelta$ on Toys and $\HpsTermShOneDelta$ on glossy Terminal. FLIP is more sensitive than SSIM to the localized specular error on Terminal (\HpsTermShOneFlip{} versus \HpsTermFlip{}).

\textbf{Reference resolution:} Increasing the reference resolution from $1280^2$ to $1808^2$ changes SSIM by $\HpsUntexRefResDelta$ on Toys and $\HpsTermRefResDelta$ on Terminal because the larger views are revisited less frequently. Textured toys benefits from $1808^2$; using $1280^2$ changes SSIM by $\HpsTexRefResDelta$.

\textbf{Input resolution:} Increasing the resolution of twenty input cameras from $256^2$ to $512^2$ changes SSIM by $+\HpsTexInResDelta$ on Textured toys, $+\HpsUntexInResDelta$ on Toys, and $\HpsTermInResDelta$ on Terminal over the same evaluation interval. The higher resolution initializes two to three times as many primitives (\HpsTexInResPrims{}k versus \HpsTexPrims{}k), which improves quality only when the additional primitives represent previously unresolved texture. Changes due to initial Gaussian coverage remain within the observed run variation.

%\rev{Available memory also constrains the reference-view count: thirty-two $1808^2$ views exceed the 12\,GB GPU capacity on Textured toys. The selected configuration must therefore balance scene coverage, per-view optimization frequency, and memory use.}

\subsection{Convergence in Static Scenes}

%\rev{All experiments in this section use the deployed camera configuration, in which the reference-camera rig follows the moving viewer. Section~\ref{sec:eval_params} evaluates the reference-view count, and Figure~\ref{fig:convergence} reports convergence under this configuration.}

\begin{figure}[t]
    \centering
    \includegraphics[width=\columnwidth]{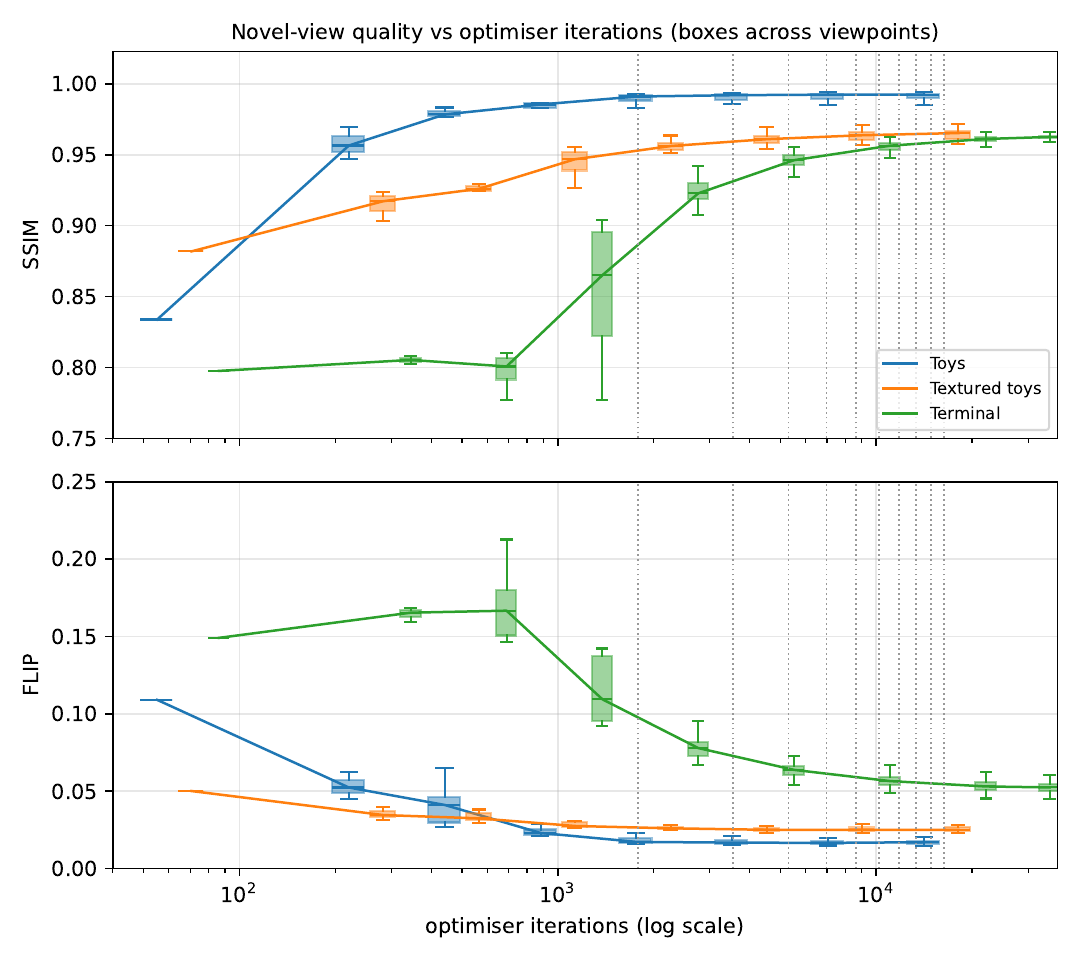}
    \caption{Convergence under the moving reference-camera rig. Boxes aggregate exponentially increasing iteration intervals. Dotted lines indicate 30-s trajectory laps (ten for Toys and twenty for Terminal).}
    \label{fig:convergence}
\end{figure}

We first measure how the visual quality of the 3D Gaussian model converges using the best parametrization for each scene found through the hyperparameter sweep. Figure~\ref{fig:convergence} reports quality as a function of optimizer iterations, while Section~\ref{sec:eval_params} separately evaluates the effect of the reference-view count.

% \begin{figure}[h]
%     \centering
%     \includegraphics[width=0.8\columnwidth]{figures/orbitview_staticref_flip.png}
%     \caption{Moving view camera with static reference cameras.}
%     \label{fig:orbitview_staticref}
% \end{figure}

% \begin{figure}[h]
%     \centering
%     \includegraphics[width=0.8\columnwidth]{figures/orbitrig_flip.png}
%     \caption{Moving the entire camera rig.}
%     \label{fig:orbitview_staticref}
% \end{figure}

Convergence depends strongly on scene content. Toys plateaus after 5-7k iterations (three to four laps) at SSIM \HpsUntexSsim{}. Textured toys, with \HpsTexOpPrims{}k primitives compared with \HpsUntexPrims{}k on Toys, plateaus after about 10k iterations (six laps) at \HpsTexOpSsim{}. Terminal converges more slowly, reaching SSIM 0.962 only after roughly 20k iterations.

%\rev{A 30\,s lap yields approximately 1700 optimizer steps on the toy scenes and 1500 on Terminal. Because Terminal contains two to six times as many primitives, it receives only 1.5--2.1 steps per lap per thousand primitives, compared with 11.5 on Toys. After three laps, Terminal remains approximately 0.019 SSIM below its converged value; quality measurements must therefore be reported together with iteration or lap count.}

A 30-s lap yields approximately 1700 optimizer steps on the toy scenes and 1500 on Terminal. Because Terminal contains two to six times as many primitives, it receives only 1.5-2.1 optimizer steps per lap per thousand primitives, compared with 11.5 on Toys. After three laps, Terminal remains approximately 0.019 SSIM below its later converged value. %This motivates reporting optimization progress alongside quality in the remaining online experiments.

\subsection{Viewpoint Robustness under Pose Offset}
\label{sec:eval_offset}

We next evaluate how the visual quality changes with camera pose and compare the 3DGS proxy to the video-based baselines (Section \ref{sec:baselines}) at eleven anchor positions. Note that these experiments exclude quality loss due to video compression to expose the geometry of latency compensation (Section~\ref{sec:streaming} includes encoded transmission).

\begin{figure}[t]
    \centering
    \includegraphics[width=\columnwidth]{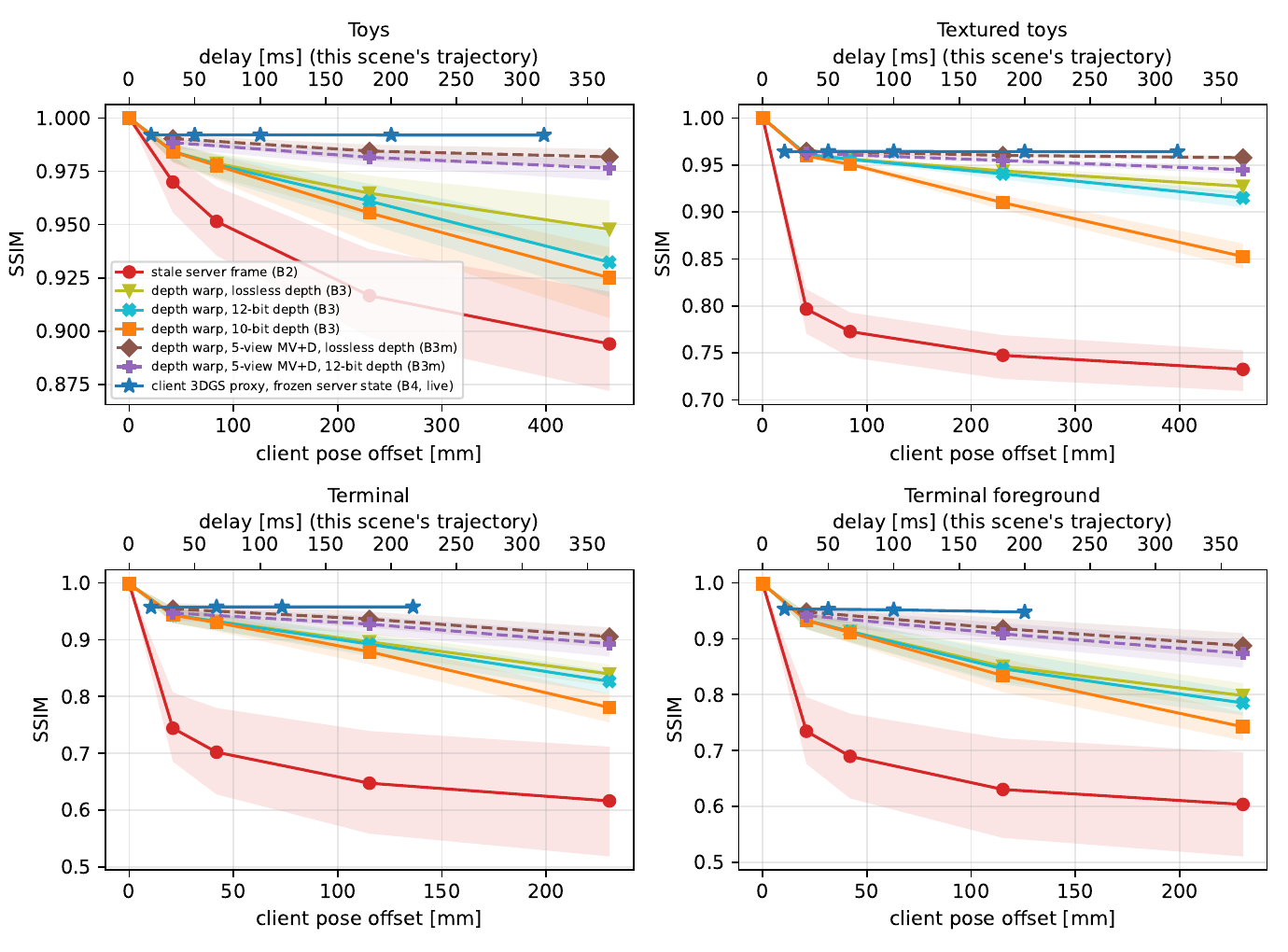}
    \caption{Rendering quality as a function of displacement from the most recent server pose (bottom axis) and the corresponding latency (top axis). Bands span the 5th--95th percentiles over eleven anchor positions.}
    \label{fig:f3_offset}
\end{figure}

\textbf{Proxy quality remains stable with pose offset:} With the server state frozen, SSIM remains \ProxyUntexSsim{} (FLIP \ProxyUntexFlip{}) on Toys, \ProxyTexSsim{} (\ProxyTexFlip{}) on Textured toys, and \ProxyTermSsim{} (\ProxyTermFlip{}) on Terminal. The quality is bounded by the content-dependent reconstruction capability of the proxy.
%$\ProxyUntexSlopePerM$, $\ProxyTexSlopePerM$, and $\ProxyTermSlopePerM$ SSIM per metre.

\textbf{Image-based quality decays with offset:} On Textured toys, a stale frame falls to SSIM \StaleTexSsimTwo{} after two frames and \StaleTexSsimTwentyTwo{} at 367\,ms. Lossless-depth warping improves these values to \WarpFloatTexSsimTwo{} and \WarpFloatTexSsimTwentyTwo{}, but leaves \WarpFloatTexHolesTwentyTwo{}\,\% of content pixels empty at the largest offset.

Depth-quantization range strongly affects accuracy. Over the 0.01--100\,m camera clip range (not shown in the figure), 12- and 8-bit inverse depth provide only \DepthCodesTwelve{} and \DepthCodesEight{} usable levels for content at 2.8--9.1\,m. At 367\,ms, the corresponding SSIM values decrease to \WarpTwelveFixTexSsimTwentyTwo{} and \WarpEightFixTexSsimTwentyTwo{}. Increasing the fixed near plane to 0.1\,m preserves the bit depth and recovers nearly all lossless-depth quality (shown in the figure).

\textbf{3D warp error components:} Disocclusions are content that should become visible when a frame is reprojected from server to client view point but is not rendered in the frame transmitted by the server. Filling holes can remedy this to some extent. Depth quantization, which is in practice a must for reasonable coding efficiency~\cite{siekkinen23tomm}, reduces warping precision and the error increases with latency. The remaining deficit is view-dependent shading plus resampling.

Table~\ref{tab:decomp} shows these different sources of error at maximum offset for 12-bit depth over 0.1--100\,m. In order to quantify the components other than disocclusions, hole filling was done using the ground truth. With this range, depth costs only \DecTexDepth{} SSIM on Textured toys and \DecTermDepth{} on Terminal, versus \DecTexClipDepth{} and \DecTermClipDepth{} over the camera clip range. Shading dominates the toy scenes (\DecUntexShading{}, \DecTexShading{}), whereas disocclusion dominates Terminal (\DecTermDisocc{}, \DecTermNearDisocc{}).

\begin{table}[t]
    \centering
    \caption{Decomposition of the distortion measured by loss of SSIM score at maximum pose offset for one-view depth-assisted warping using 12-bit depth.}
    \label{tab:decomp}
    \small
    \resizebox{\columnwidth}{!}{\begin{tabular}{lrrrrrr}
\toprule
scene & offset & depth & shading & disocc. & total & holes \\
 & (mm) & \multicolumn{4}{c}{SSIM loss} & (\%) \\
\midrule
Toys & \DecUntexMm & \DecUntexDepth & \DecUntexShading & \DecUntexDisocc & \DecUntexTotal & \DecUntexHoles \\
Textured toys & \DecTexMm & \DecTexDepth & \DecTexShading & \DecTexDisocc & \DecTexTotal & \DecTexHoles \\
Terminal & \DecTermMm & \DecTermDepth & \DecTermShading & \DecTermDisocc & \DecTermTotal & \DecTermHoles \\
Terminal foreground & \DecTermNearMm & \DecTermNearDepth & \DecTermNearShading & \DecTermNearDisocc & \DecTermNearTotal & \DecTermNearHoles \\
\bottomrule
\end{tabular}
}
\end{table}

%To further quantify the effect of view-dependent shading,  With lossless depth and ground-truth hole filling, the centered Textured-toys crop falls from 1.0000 at zero offset to \DecTexCentreOne{} at \DecTexMm{}\,mm; five views reach only \DecTexCentreFive{} because all carry shading from the same server pose. At the deployable 50-Mbit/s operating point, the Terminal proxy scores \ProxyTermCentre{} on the crop and \ProxyTermFull{} on the full frame, versus \RdTermCentreThree{} and \RdTermFullThree{} for the best affordable warp. This comparison identifies an irreducible warp error; it does not claim that a finite Gaussian model always beats an uncompressed, hole-filled warp.

\textbf{Near geometry increases disocclusion:} Adding 10\,cm props at 0.45\,m (i.e., Terminal foreground vs. Terminal scene) raises holes from \DecTermHoles{}\,\% to \DecTermNearHoles{}\,\% and the disocclusion deficit from \DecTermDisocc{} to \DecTermNearDisocc{}. The proxy loses 0.005 SSIM, versus 0.041 for one-view warping. Shading error also rises from \DecTermShading{} to \DecTermNearShading{} because the props are glossy.

\textbf{Additional views remedy disocclusions but may increase depth-induced error:} With lossless depth on Toys, three and five views reduce the unfilled-pixel fraction from \WarpFloatUntexHolesTwentyTwo{}\,\% to \WarpThreeViewUntexHolesTwentyTwo{}\,\% and \WarpFiveViewUntexHolesTwentyTwo{}\,\%, increasing SSIM from \WarpFloatUntexSsimTwentyTwo{} to \WarpThreeViewUntexSsimTwentyTwo{} and \WarpFiveViewUntexSsimTwentyTwo{}. With 12-bit depth, independently quantized surfaces compete in the merged z-buffer. On Terminal at maximum offset, the one-, three-, and five-view configurations attain \WarpTwelveFixTermSsimTwentyTwo{}, \WarpThreeViewTwelveTermSsimTwentyTwo{}, and \WarpFiveViewTwelveTermSsimTwentyTwo{}, respectively. Three views improve quality when disocclusion is the dominant error, whereas five views improve neither scene. Perfect prediction of the future server viewpoint would reduce the residual offset represented on the horizontal axis of Figure~\ref{fig:f3_offset}.

\textbf{Relative performance depends on scene content:} The visual quality of 3DGS proxy is either on par (shortest latencies) or outperforms all the baselines on all the scenes. However, the quality of the baselines decreases more dramatically with latency on the Terminal scenes.
%On Toys, the proxy exceeds all image-based configurations. On Textured toys, lossless warping and the proxy remain within 0.006 SSIM.
%one-view warping attains \WarpFloatTexSsimTwo{} at \OffsetMmTexTwo{}\,mm, five-view warping attains \WarpFiveViewTexSsimTwentyTwo{} at \OffsetMmTexTwentyTwo{}\,mm, and the proxy remains at \ProxyTexSsim{}. 
%On Terminal, the stale-frame baseline clearly decreases from \StaleTermSsimTwo{} to \StaleTermSsimTwentyTwo{}, lossless one-view warping decreases from \WarpFloatTermSsimTwo{} to \WarpFloatTermSsimTwentyTwo{}, and five-view warping attains \WarpFiveViewTermSsimTwentyTwo{} at maximum offset; all remain below the proxy value of \ProxyTermSsim{}.

\subsection{Progressive Model Expansion}

A central feature of the proposed system is that the Gaussian model is not only
optimized but also expanded online as the viewer enters previously unseen
regions. We therefore evaluate the effectiveness of the Gaussian insertion
strategy separately from optimization of an already existing model.
We perform a scene exploration experiment in which we duplicate the Toys scene content to make a larger scene and then walk around it and track system behavior through GPU memory use, number of total and frozen 3D Gaussians, and computational performance of the optimization.

\begin{figure}[h]
    \centering
    \includegraphics[width=\columnwidth]{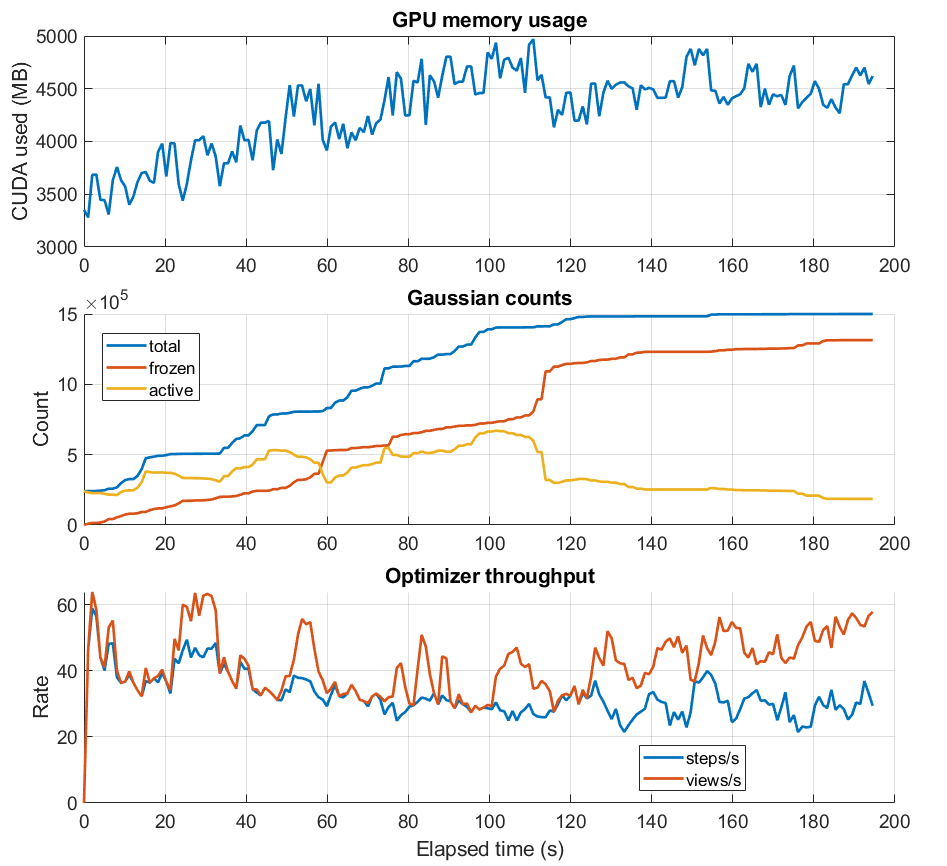}
    \caption{System metrics during the exploration experiment.}
    \label{fig:drive_through}
\end{figure}

The results are plotted in Figure \ref{fig:drive_through}. They highlight a few important aspects: First, while the CUDA memory use grows with the number of added Gaussians, it remains reasonable in part thanks to the freezing mechanism. Second, freezing also alleviates the optimizer workload as we can see from elapsed time 150 onwards (views/s being larger than steps/s means optimizer is batching). A modern desktop graphics card typically has at least 12GB of VRAM available, so we can estimate that this kind of a system should handle at least a few million Gaussians and servers with datacenter grade GPUs much more than that.

\subsection{Object Motion Tracking and Relighting}

We next evaluate the quality when there are moving objects and dynamic lighting in the scene. We add a bouncing textured ball and a rotating white cube into Toys and rotate 15 degrees per second the single directional light that the scene has at 45 degree zenith angle. We either run the system without object tracking and relighting or enable both. The reason is that real-time shadow mapping is needed when objects move even if the light itself does not move.

\begin{figure}[h]
    \centering
    \includegraphics[width=0.8\columnwidth]{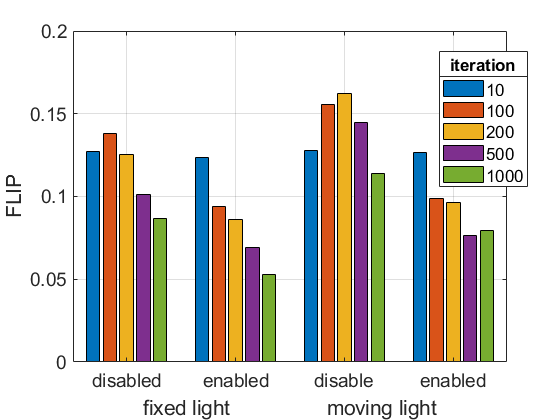}
    \caption{View synthesis quality under scene dynamics.}
    \label{fig:motion_relight}
\end{figure}

\begin{figure}[ht]
    \centering
    \begin{subfigure}[b]{0.15\textwidth}
        \centering
        \includegraphics[width=\linewidth]{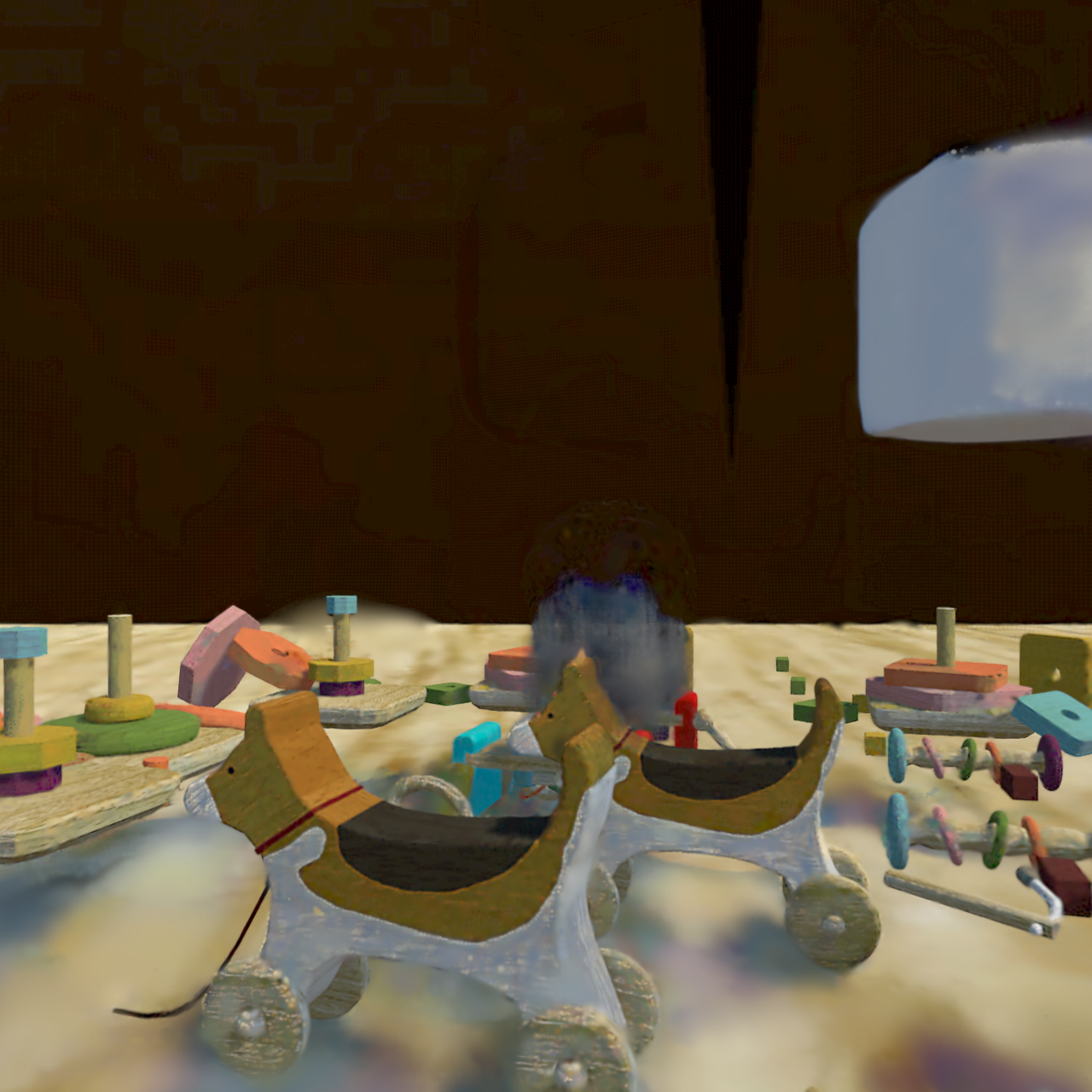}
        \caption{Disabled}
        \label{fig:example_norelight}
    \end{subfigure}
    \hfill
    \begin{subfigure}[b]{0.15\textwidth}
        \centering
        \includegraphics[width=\linewidth]{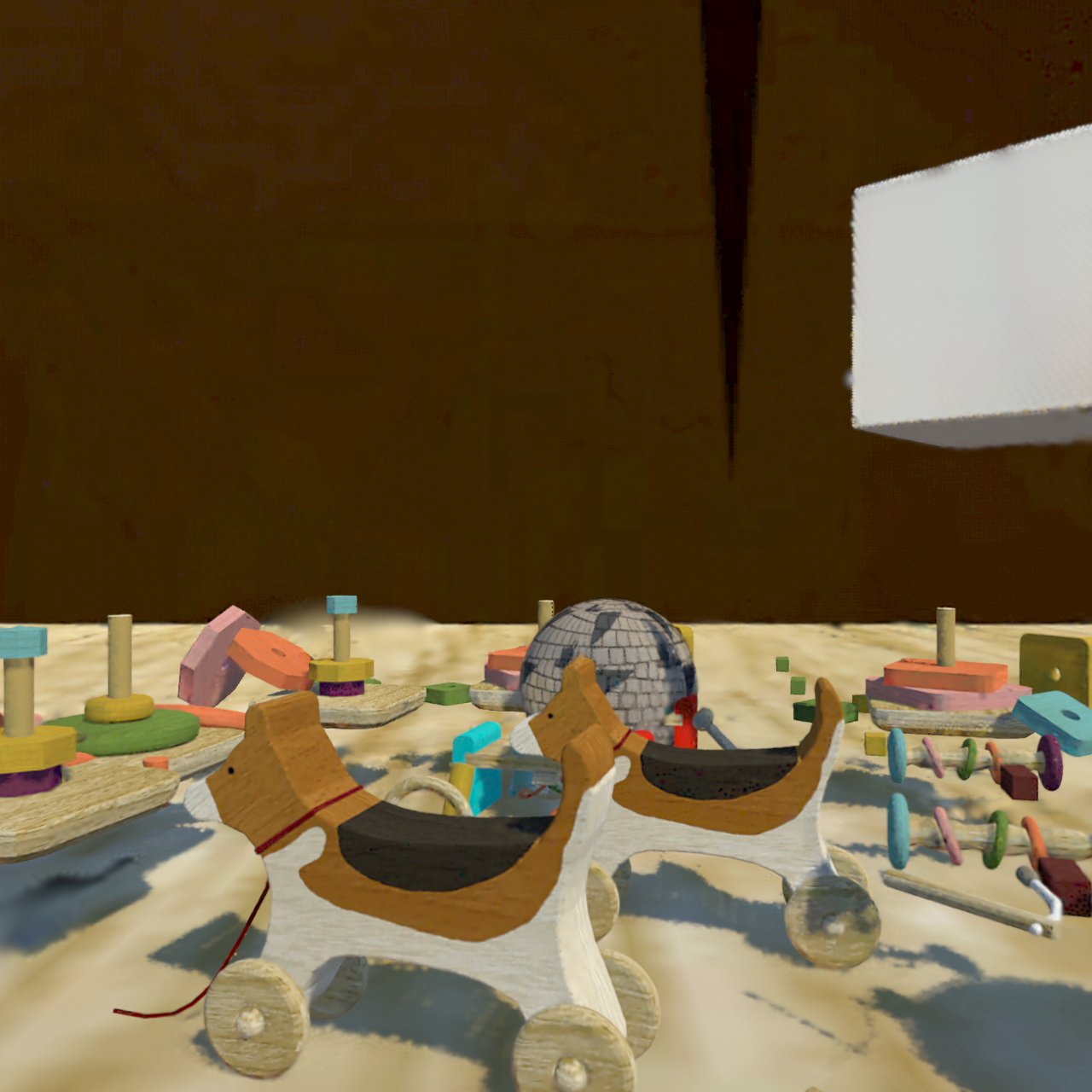}
        \caption{Enabled}
        \label{fig:example_relight}
    \end{subfigure}
    \begin{subfigure}[b]{0.15\textwidth}
        \centering
        \includegraphics[width=\linewidth]{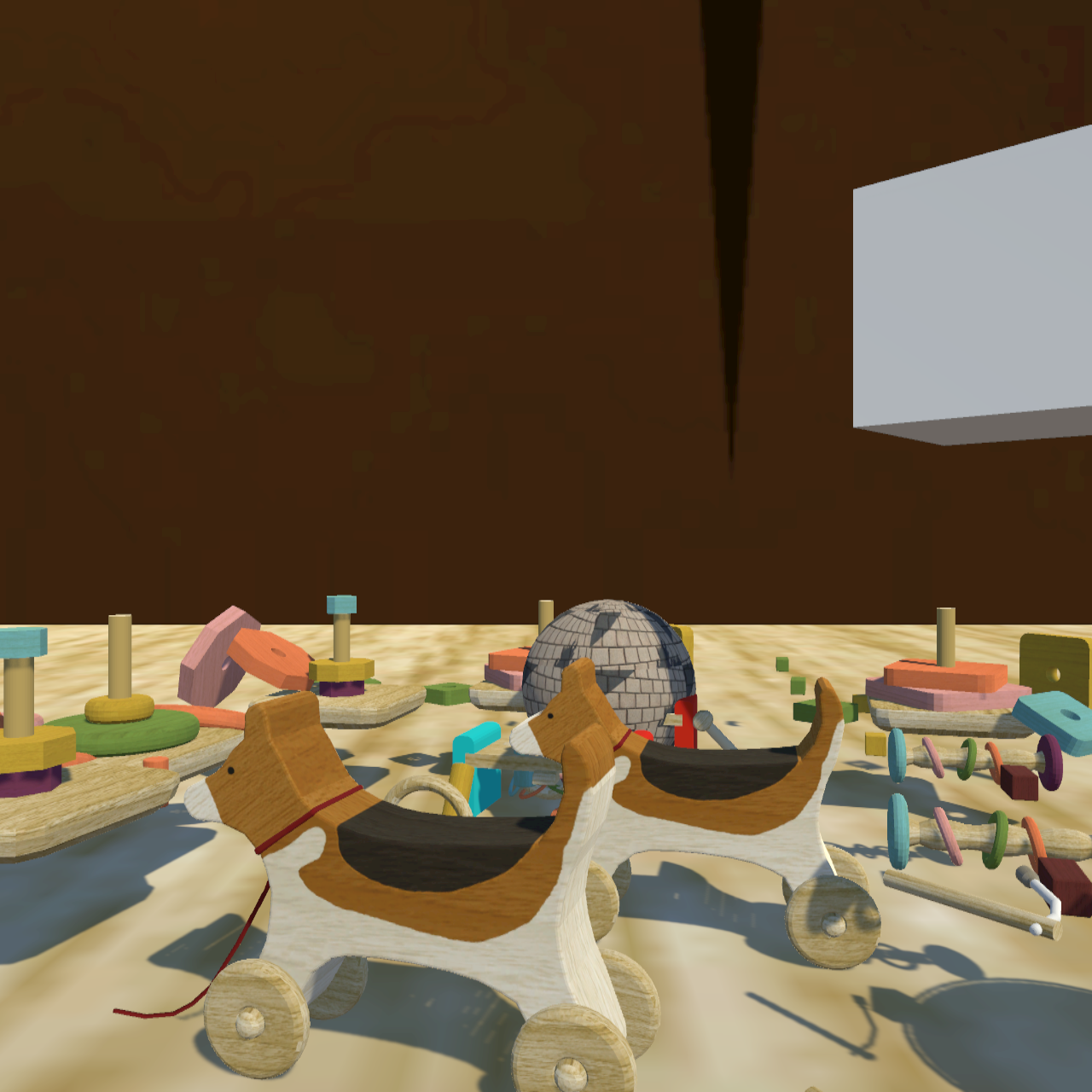}
        \caption{Ground truth}
        \label{fig:example_gt}
    \end{subfigure}
    \caption{Rendered frames from 3D Gaussian model with and without support for dynamics.}
    \label{fig:samples}
\end{figure}

The results shown in Figure \ref{fig:motion_relight} demonstrate that explicit support for relighting and object tracking dramatically improves the quality when the scene is dynamic. We show sample screenshots in Figure \ref{fig:samples} where object and light motion blur the model during optimization when not explicitly accounted for.

%\subsection{Geometric Integrity of Object-Bound Gaussians}
%\label{sec:eval_integrity}

Gaussians bound to a rigid object are optimized in image space through the object's tracked transform (Section~\ref{sec:relight_dyn}). If that transform lagged behind, or differed from, the object pose at which a reference image was rendered, the image-space gradients would displace the Gaussians toward where the object appeared in the reference images, smearing them along the motion path and onto the background (Figure~\ref{fig:example_norelight} shows this failure qualitatively). We test whether it occurs with tracking enabled by optimizing the dynamic scene with the bouncing ball (true radius 2.00\,m) for 2000 frame-paced frames (33\,s at 60\,fps) with tracking and relighting enabled, and saving the final model with its per-Gaussian object IDs.

Table~\ref{tab:f5_integrity} characterizes the Gaussians bound to the ball by the distance of their centers from the center of a least-squares sphere fitted to them. The results in the table demonstrate that the centers form a shell of a few centimeters just inside the surface. No center lies beyond $1.25\times$ the true radius. Applying the object-local transform before every optimization step therefore keeps rigidly bound Gaussians spatially coherent over 2000 frames of motion.

\begin{table}[t]
    \centering
    \caption{Spatial integrity of the Gaussians bound to the bouncing ball after 2000 optimization frames: least-squares sphere fit to the Gaussian centres, distance of the centres from the fitted centre, and the fraction of centres beyond $1.25\times$ and $2\times$ the true radius.}
    \label{tab:f5_integrity}
    \small\setlength{\tabcolsep}{4pt}
    \begin{tabular}{lr}
\toprule
quantity & ball, 122\,223 Gaussians \\
\midrule
fitted / true radius (m) & 1.97 / 2.00 \\
radial distance p50 / p95 / max (m) & 1.97 / 1.98 / 2.00 \\
mean absolute radial deviation (m) & 0.03 \\
centres beyond $1.25\times$ radius & 0.00\,\% \\
\bottomrule
\end{tabular}

\end{table}

\section{Streaming}
\label{sec:streaming}
We evaluate the end-to-end streaming system under rate constraints by comparing the client-rendered Gaussian proxy with one-, three-, and five-view depth-assisted warping at matched rate ceilings, with controlled server-view delays for warping. All image-based configurations are evaluated against Unity ground truth rendered at the matching client pose.

\begin{table*}[t]
    \centering
    \caption{Depth-assisted warping and 3DGS proxy quality at different target bitrates and latencies. Italic entries denote configurations in which lossless depth exceeds the rate ceiling and rate-controlled depth is used instead. Bold entries denote the best value in each column, each half of a pair separately.}
    \label{tab:wt}
    \small\setlength{\tabcolsep}{3pt}
    \resizebox{\textwidth}{!}{

\begin{tabular}{ll rrr rrr rrr rrr rrr}
\toprule
 & & \multicolumn{12}{c}{SSIM at 67 / 183 / 367\,ms of delay, by total rate (Mbit/s)} & \multicolumn{3}{c}{367\,ms, at 50 / 100\,Mbit/s} \\
\cmidrule(lr){3-14} \cmidrule(lr){15-17}
scene & configuration & \multicolumn{3}{c}{10} & \multicolumn{3}{c}{25} & \multicolumn{3}{c}{50} & \multicolumn{3}{c}{100} & FLIP & holes & TRD \\
\cmidrule(lr){3-5} \cmidrule(lr){6-8} \cmidrule(lr){9-11} \cmidrule(lr){12-14}
Toys & 1 view & \textbf{\textit{0.958}} & \textit{0.914} & \textit{0.851} & 0.961 & 0.918 & 0.856 & 0.962 & 0.919 & 0.857 & 0.962 & 0.919 & 0.857 & 0.104 / 0.103 & 16.2\% / 16.2\% & 0.0177 / 0.0177 \\
 & 3 views & \textit{0.936} & \textit{0.917} & \textit{0.885} & \textit{0.963} & \textit{0.941} & \textit{0.908} & \textbf{0.972} & 0.949 & 0.915 & 0.973 & 0.950 & 0.916 & 0.076 / 0.075 & 12.3\% / 12.3\% & 0.0110 / 0.0108 \\
 & 5 views & \textit{0.905} & \textit{0.893} & \textit{0.868} & \textit{0.940} & \textit{0.928} & \textit{0.901} & \textit{0.964} & \textit{0.950} & \textit{0.923} & \textbf{0.979} & 0.963 & 0.934 & \textit{0.064} / 0.060 & \textit{10.4\%} / 10.4\% & \textit{0.0109} / 0.0097 \\
 & 3DGS proxy, streamed & 0.953 & \textbf{0.953} & \textbf{0.953} & \textbf{0.964} & \textbf{0.964} & \textbf{0.964} & 0.968 & \textbf{0.968} & \textbf{0.968} & 0.969 & \textbf{0.969} & \textbf{0.969} & \textbf{0.048} / \textbf{0.046} & \textbf{0.003\%} / \textbf{0.003\%} & \textbf{0.0066} / \textbf{0.0065} \\
\midrule
Textured toys & 1 view & \textbf{\textit{0.892}} & \textbf{\textit{0.865}} & \textit{0.822} & \textbf{0.923} & \textbf{0.892} & 0.847 & \textbf{0.933} & 0.899 & 0.853 & \textbf{0.937} & 0.902 & 0.855 & 0.070 / 0.068 & 16.2\% / 16.2\% & 0.0199 / 0.0196 \\
 & 3 views & \textit{0.827} & \textit{0.816} & \textit{0.792} & \textit{0.887} & \textit{0.874} & \textit{0.848} & 0.917 & \textbf{0.901} & 0.873 & 0.936 & 0.917 & 0.887 & 0.060 / 0.056 & 12.3\% / 12.3\% & 0.0180 / 0.0170 \\
 & 5 views & \textit{0.777} & \textit{0.771} & \textit{0.753} & \textit{0.840} & \textit{0.833} & \textit{0.812} & \textit{0.894} & \textit{0.886} & \textit{0.864} & 0.930 & \textbf{0.918} & \textbf{0.894} & \textbf{\textit{0.055}} / \textbf{0.048} & \textit{10.4\%} / 10.4\% & \textit{0.0187} / \textbf{0.0164} \\
 & 3DGS proxy, streamed & 0.844 & 0.844 & \textbf{0.844} & 0.867 & 0.867 & \textbf{0.867} & 0.876 & 0.876 & \textbf{0.876} & 0.878 & 0.878 & 0.878 & 0.092 / 0.089 & \textbf{0.150\%} / \textbf{0.156\%} & \textbf{0.0168} / 0.0167 \\
\midrule
Terminal & 1 view & \textbf{0.879} & 0.817 & 0.722 & 0.902 & 0.837 & 0.739 & 0.910 & 0.844 & 0.744 & 0.914 & 0.848 & 0.747 & 0.168 / 0.165 & 12.7\% / 12.7\% & 0.0278 / 0.0275 \\
 & 3 views & \textit{0.834} & \textit{0.813} & \textit{0.773} & 0.889 & 0.865 & 0.819 & \textbf{0.913} & 0.888 & 0.841 & \textbf{0.926} & \textbf{0.899} & 0.851 & 0.127 / 0.122 & 7.1\% / 7.1\% & 0.0187 / 0.0177 \\
 & 5 views & \textit{0.799} & \textit{0.782} & \textit{0.749} & \textit{0.831} & \textit{0.814} & \textit{0.781} & 0.898 & 0.878 & 0.839 & 0.918 & 0.897 & 0.858 & \textbf{0.121} / \textbf{0.113} & 5.5\% / 5.5\% & 0.0196 / 0.0181 \\
 & 3DGS proxy, streamed & 0.870 & \textbf{0.870} & \textbf{0.870} & \textbf{0.904} & \textbf{0.904} & \textbf{0.904} & 0.904 & \textbf{0.904} & \textbf{0.904} & 0.890 & 0.890 & \textbf{0.890} & 0.128 / 0.132 & \textbf{0.172\%} / \textbf{1.804\%} & \textbf{0.0164} / \textbf{0.0171} \\
\midrule
Terminal foreground & 1 view & \textit{0.843} & \textit{0.779} & \textit{0.675} & 0.882 & 0.816 & 0.709 & \textbf{0.897} & 0.829 & 0.720 & 0.904 & 0.836 & 0.726 & 0.175 / 0.171 & 10.5\% / 10.5\% & 0.0355 / 0.0349 \\
 & 3 views & \textit{0.791} & \textit{0.769} & \textit{0.741} & \textit{0.842} & \textit{0.818} & \textit{0.787} & 0.880 & 0.855 & 0.821 & \textbf{0.910} & 0.884 & 0.848 & 0.129 / 0.118 & 3.6\% / 3.6\% & 0.0259 / 0.0239 \\
 & 5 views & \textit{0.747} & \textit{0.731} & \textit{0.707} & \textit{0.795} & \textit{0.778} & \textit{0.752} & 0.805 & 0.788 & 0.762 & 0.894 & 0.875 & 0.845 & 0.148 / \textbf{0.116} & 2.7\% / 2.7\% & 0.0295 / 0.0233 \\
 & 3DGS proxy, streamed & \textbf{0.863} & \textbf{0.863} & \textbf{0.863} & \textbf{0.896} & \textbf{0.896} & \textbf{0.896} & 0.890 & \textbf{0.890} & \textbf{0.890} & 0.888 & \textbf{0.888} & \textbf{0.888} & \textbf{0.126} / 0.119 & \textbf{0.143\%} / \textbf{0.381\%} & \textbf{0.0203} / \textbf{0.0210} \\
\bottomrule
\end{tabular}

}
\end{table*}

\subsection{Streaming Evaluation Protocol}
\label{sec:streaming_bw}

\textbf{Evaluation intervals:} For Toys, Textured toys, and Terminal, we select three 3-s intervals from the 60-s walkthrough representing the fastest translation, slowest translation, and largest rotation. For Terminal foreground, we instead use a single 3-s interval covering the closest approach to the added near-field objects. All configurations use the same client-pose trace and evaluation windows at delays of 67, 183, and 367~ms.

\textbf{Streamed proxy:}
Each run uses twenty input cameras, sixteen reference cameras that follow the current client pose, degree-three SH, and Zstd block compression. Toys uses $256^2$ input and $1280^2$ reference images with 1.0-pixel Gaussian coverage. Textured toys uses $512^2$ input and $1808^2$ reference images with the same coverage. Both Terminal variants use $256^2$ input and $1280^2$ reference images with 2.5-pixel coverage.

Before the measured walkthrough, the server repeats an orbit with a 30-s period until the optimizer has completed at least 5k additional steps on Toys, 10k on Textured toys, or 15k on either Terminal scene.
Opacity resets and joint densification and pruning are then disabled to preserve stable Gaussian identities during streaming. Gradient optimization, client-conditioned acquisition, tile selection, and streaming continue throughout the scored walkthrough.

\textbf{Sampling and rate control:}
Quality is scored at a nominal stride of two trace frames within each 3-s interval, giving 90 intended images per interval. Metrics are averaged within each interval and then equally across intervals. For each sample, a native marker records the pose actually used by the client renderer. %We accept a run only when the recorded pose, trajectory identifier, camera intrinsics, image dimensions, and protocol accounting match the predefined configuration.
Client output is compared directly with backend-matched Unity ground truth rendered at the same pose, both at $1280\times1280$ pixels. Table~\ref{tab:wt} uses LDR-FLIP with its default viewing conditions, approximately 67 pixels per degree. %, without image registration or transform search.
We evaluate target rates of 10, 25, 50, and 100~Mbit/s using a 512-KiB burst allowance and a 100-ms model adoption interval. Reported rates are measured over the full 60-s walkthrough and include all downstream TCP application-protocol traffic, including framing and control messages. The Toys runs exceed their targets by only 0.09--0.10~Mbit/s. The 100~Mbit/s Terminal and Terminal-foreground runs are source- and client-limited at 85.87 and 85.64~Mbit/s, respectively. Each rate point comes from a separate continuously optimized run, so small non-monotonic differences across rates may include run-to-run variation.

\textbf{Depth-assisted warping:}
Each reference view is transmitted as a low-latency NVENC HEVC colour stream together with a monochrome depth stream. The client unprojects the decoded pixels and forward-warps them to the current client pose using a z-buffered point splat. Single-pixel cracks are filled, while disoccluded regions remain unfilled. The one-view configuration uses the server-centred view. The three-view configuration additionally uses views translated \MvBaselineMm\,mm to the left and right, while the five-view configuration also adds views above and below. All views are merged into a common client z-buffer. During the imposed delay, the reference-view rig remains at the most recently received server pose.

Colour is encoded with NVENC HEVC using preset p1, CBR, a one-frame buffer, GOP 60, and no B-frames. Depth uses 12-bit inverse depth over 0.1--100\,m and is encoded with x265. Because warp quality is highly sensitive to depth precision, lossless depth is used whenever it fits within the target budget and depth allocation is prioritized before the remaining rate is assigned to colour. When lossless depth alone exceeds the budget, the depth stream is rate controlled with x265 using a one-frame buffer and intra refresh, with 70\% of the per-view allocation reserved for depth. Because the views are encoded independently rather than with MV-HEVC, the reported multi-view rates are conservative upper bounds.

\textbf{Additional metrics:}
In addition to the SSIM and FLIP metrics defined in Section~\ref{sec:evaluation}, we report the unfilled-pixel fraction and the temporal residual difference $\mathrm{TRD}=\mathrm{mean}\,|e_t-e_{t-2}|$, where $t$ indexes the 60-Hz pose trace and $e_t$ is the RGB residual relative to ground truth. TRD measures frame-to-frame instability in reconstruction error and is therefore sensitive to temporal artefacts such as flicker, popping, unstable hole filling, and inconsistent reprojection or silhouette motion. For the proxy, TRD is computed only from exact two-frame-stride pairs within the same 3-s evaluation interval. Proxy holes are pixels whose ground-truth maximum RGB exceeds $2/255$ while the client maximum RGB is at most $2/255$, expressed as a fraction of the full image. Warping holes are invalid pixels in its raster-validity mask.

%\rev{\textbf{Metrics.} In addition to SSIM and FLIP, we report the unfilled-pixel fraction and the temporal residual difference $\mathrm{TRD}=\mathrm{mean}\,|e_t-e_{t-1}|$ at a two-frame stride, where $e_t$ is the residual relative to ground truth at frame $t$; for the proxy, TRD uses only exact two-frame-stride pairs from the same 3-s intervals. As a reference for the coding cost alone, low-latency HEVC of the same $1280^2$ frames at the server pose, without pose offset, attains SSIM \HevcTwentyFiveSsim{} at \HevcTwentyFiveMbps~Mbit/s and \HevcFiftySsim{} at \HevcFiftyMbps~Mbit/s on the Toys orbit of Section~\ref{sec:quality}, above the proxy values of \ProxyUntexSsim{} (Toys) and \ProxyTexSsim{} (Textured toys) measured there; the comparison below concerns the client pose.}

\subsection{Results}
\label{sec:eval_wt}

\begin{figure*}[t]
    \centering
    \includegraphics[width=\textwidth]{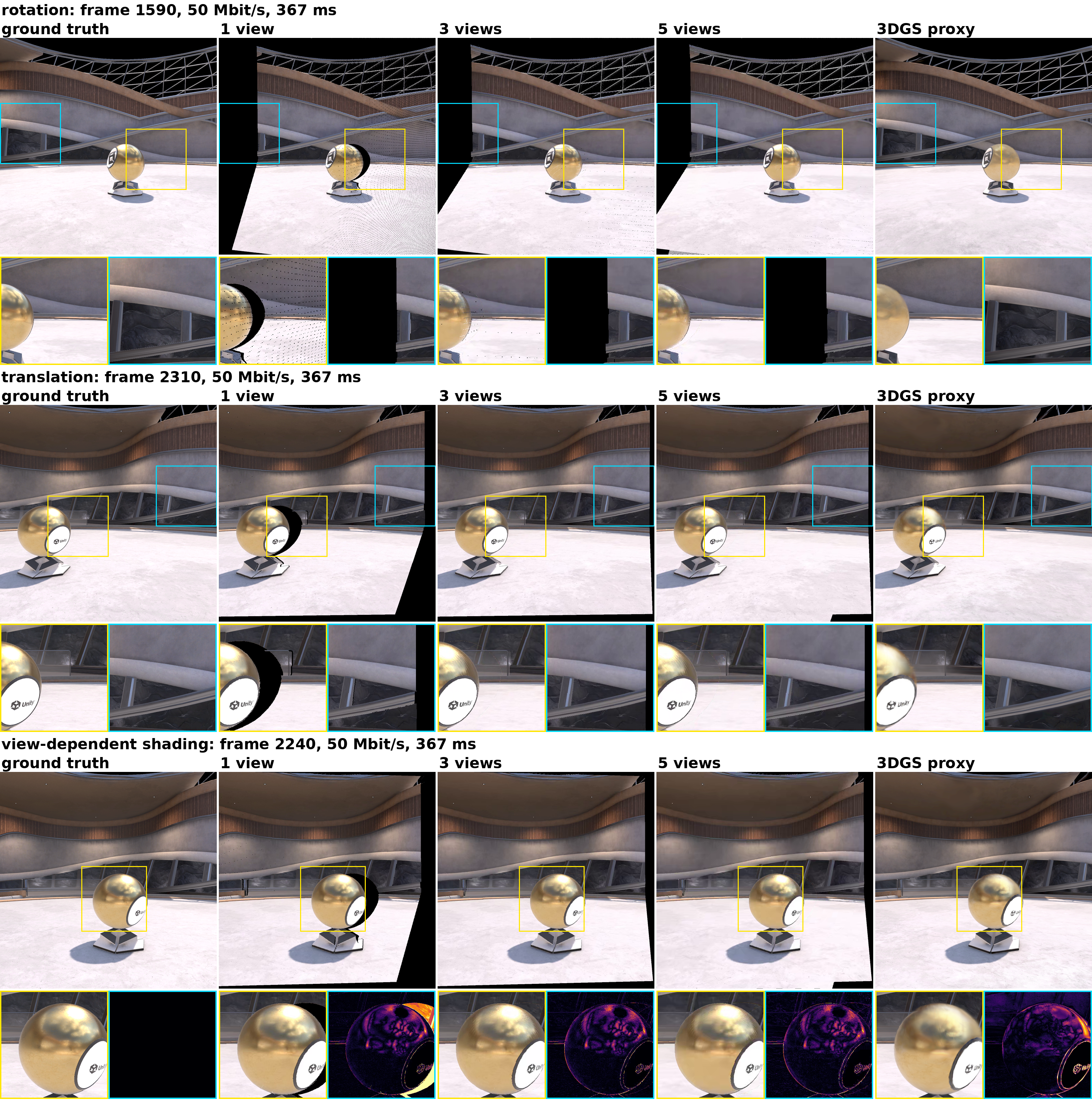}
   \caption{Client frames on the Terminal walkthrough at 50~Mbit/s and 367\,ms of delay: rotation (frame 1590), translation (2310), and view-dependent shading near the sphere (2240). Insets magnify selected regions. In the last row, cyan-bordered insets show mean absolute RGB error against ground truth on a shared 0--0.3 scale. Brighter colours indicate larger errors.}
    \label{fig:wt_samples}
\end{figure*}

Table~\ref{tab:wt} reports SSIM for all tested rate ceilings and delays, together with FLIP, unfilled-pixel fraction, and TRD at 367~ms for 50 and 100~Mbit/s. Because the proxy is rendered from the current client pose rather than from a delayed server viewpoint, its quality does not depend on the imposed delay in this experiment. The same per-rate proxy value is therefore shown in each delay column.

\textbf{The proxy becomes advantageous as pose offset increases:}
At 50~Mbit/s, depth-assisted warping attains the highest SSIM at 67~ms on all four scenes. The streamed proxy exceeds all evaluated warping configurations at 183~ms on Toys, Terminal, and Terminal foreground, and at 367~ms on all four scenes. The crossover therefore lies between 67 and 183~ms for three scenes, and between 183 and 367~ms for Textured toys. On Terminal, the proxy attains SSIM 0.904, compared with 0.913, 0.888, and 0.841 for the strongest warping configuration at 67, 183, and 367~ms, respectively.

The crossover depends on both scene and rate. At 100~Mbit/s, the proxy is ahead at 183~ms on Toys and Terminal foreground, while Terminal crosses over between 183 and 367~ms. On Textured toys, five-view warping remains ahead even at 367~ms (SSIM 0.894 versus 0.878 for the proxy). Table~\ref{tab:wt} reports the complete per-scene results.

\textbf{Rate affects the two representations differently:} Increasing the target rate from 10 to 50~Mbit/s improves proxy SSIM by 0.016--0.034 across the four scenes. Raising the ceiling to 100~Mbit/s produces little additional improvement on the toy scenes and lower SSIM in both Terminal runs. Each rate point comes from a separate continuously optimized run. %Without repeated trials, the significance of these non-monotonic differences cannot be established.

%For warping, depth precision is the dominant rate constraint. At the \MvBaselineMm\,mm multi-view baseline, using 12-bit inverse depth over 0.1--100\,m reduces the image-space error of one quantization step from 12.2 to 1.2 pixels compared with the full 0.01--100\,m camera range. On Textured toys at maximum offset, one-view SSIM increases from 0.7552 with a 0.18-Mbit/s depth stream to 0.9071 when depth is transmitted losslessly at 1.57~Mbit/s, while increasing the colour rate by 90~Mbit/s changes SSIM by only $-0.007$. Once depth is adequately provisioned, additional reference views rather than additional colour bitrate provide most of the remaining quality improvement.

\textbf{Additional views improve spatial coverage:}
At 100~Mbit/s and 367~ms on Terminal, increasing from one to five reference views reduces the unfilled fraction from \WtTermOneKTwentyTwoHoles\,\% to \WtTermFiveKTwentyTwoHoles\,\% and increases SSIM from \WtTermOneKTwentyTwoSsim{} to \WtTermFiveKTwentyTwoSsim{}. On Toys, the corresponding SSIM increases from \WtToysOneKTwentyTwoSsim{} to \WtToysFiveKTwentyTwoSsim{}. Most of the improvement comes from increasing the view count from one to three. The fifth view provides a smaller gain and can reduce quality at 10 and 25~Mbit/s because the additional depth streams consume bandwidth that would otherwise be available for colour.

Figure~\ref{fig:wt_samples} shows the remaining limitations at 50~Mbit/s. Additional views fill disoccluded regions behind foreground objects, but the transported colour still represents the server viewpoint and therefore retains stale view-dependent shading. At the reported 50 and 100~Mbit/s rates, the proxy's black-gap fraction ranges from 0.003\,\% to 1.804\,\%, while warping's invalid-pixel fraction ranges from 2.7\,\% to 16.2\,\%.

%\rev{\textbf{Rotation produces larger errors than slow translation.} A \WtRotToysRotDeg$^\circ$ rotation over 367\,ms leaves \WtRotToysOneHoles\,\% of the frame unfilled with one view and \WtRotToysFiveHoles\,\% with five views, with SSIM values of \WtRotToysOneSsim{} and \WtRotToysFiveSsim{}, respectively, whereas the slowest-translation interval exceeds 0.95 SSIM with one view. Rotation exposes directions outside the forward-facing rig and increases the lossless-depth rate to \WtDepthMbpsMin{}--\WtDepthMbpsMax~Mbit/s per view; with lossless depth, the five-view configuration therefore does not fit within 25~Mbit/s on any scene (italic entries in Table~\ref{tab:wt}). The proxy, which renders the current pose from a persistent model, is unaffected by the rotation.}

\textbf{Rotation is particularly challenging for viewpoint-specific warping:} On Toys at 367~ms, a \WtRotToysRotDeg$^\circ$ rotation leaves \WtRotToysOneHoles\,\% of the frame unfilled with one reference view and \WtRotToysFiveHoles\,\% with five views, with SSIM values of \WtRotToysOneSsim{} and \WtRotToysFiveSsim{}, respectively. By comparison, the slowest-translation interval exceeds 0.95 SSIM with one view. Rotation exposes directions outside the forward-facing reference rig and increases the lossless-depth requirement to \WtDepthMbpsMin{}--\WtDepthMbpsMax~Mbit/s per view. As a result, the five-view configuration with lossless depth does not fit within 25~Mbit/s on any scene. The persistent proxy does not incur the same delay-induced rotation penalty because it renders the received 3D representation from the current client orientation.

%\textbf{SSIM and FLIP emphasize different error modes.}~At 100~Mbit/s and 367~ms, SSIM favours the proxy on all four scenes, whereas FLIP favours warping on Textured toys, Terminal, and Terminal foreground. Warping preserves more of the transmitted local texture and colour, but leaves disoccluded regions and retains appearance from the server viewpoint. The proxy provides more complete spatial coverage, but its appearance is limited by reconstruction error. The two metrics nevertheless agree on the rate ordering within each configuration.

\textbf{SSIM and FLIP emphasize different error modes:}
At 100~Mbit/s and 367~ms, both metrics favour the proxy on Toys and warping on Textured toys. On Terminal and Terminal foreground, SSIM favours the proxy whereas FLIP favours warping. Warping preserves transmitted local texture and colour but leaves disoccluded regions and retains appearance from the server viewpoint. The proxy's appearance remains limited by reconstruction error.

%\rev{\textbf{Temporal stability is comparable, and the proxy's residual is caused by continued optimization.} At 367~ms and 100~Mbit/s, the proxy's TRD is lower than that of five-view warping on Toys (\WtProxyToysRHundredTrd{} versus \WtToysFiveKTwentyTwoTrd{}) and Terminal foreground (\WtProxyNearRHundredTrd{} versus \WtNearFiveKTwentyTwoTrd{}) and equal within 0.001 on Textured toys and Terminal (\WtProxyTexRHundredTrd{} versus \WtTexFiveKTwentyTwoTrd{}; \WtProxyTermRHundredTrd{} versus \WtTermThreeKTwentyTwoTrd{}). The warping residual changes with the disocclusion pattern from frame to frame; the proxy residual changes because the model itself is updated between frames. In a separate Terminal run on the Section~\ref{sec:quality} orbit with dense sampling, TRD at the same stride is 0.0203 while optimization continues and 0.0135 after the server state is frozen. Converging the model before the measured pass, or applying update dead zones, therefore removes a component of the proxy's temporal error that has no counterpart in warping.}

\textbf{Temporal residual error generally favours the proxy:}
At 367~ms, the proxy has lower TRD than every evaluated warping configuration on all four scenes at 50~Mbit/s and on three scenes at 100~Mbit/s. The exception is Textured toys at 100~Mbit/s, where five-view warping attains 0.0164 versus 0.0167 for the proxy. The temporal residuals change as disocclusion patterns move across frames when warping. The proxy's continued model optimization also contributes to temporal variation.

A separate Terminal run confirms this effect: freezing the server state reduces TRD from 0.0203 to 0.0135. This indicates that some of the measured proxy variation is caused by ongoing model updates rather than by viewpoint motion itself.

%In summary, at 50~Mbit/s warping attains higher SSIM at 67~ms, whereas the proxy attains higher SSIM and lower TRD on all four scenes at 367~ms. Rendering the received representation from the current client pose avoids the delay-induced viewpoint penalty in this experiment.

\section{Discussion and Limitations}
\label{sec:discussion}

The results show that a persistent client-renderable representation changes how remote rendering responds to pose offset. Image-based reprojection retains high fidelity at small offsets, but its quality degrades as disocclusions and stale view-dependent appearance accumulate. The Gaussian proxy is instead rendered from the current client pose, so its quality is largely determined by reconstruction and synchronization quality rather than by the displacement from the latest server-rendered view. This produces a clear operating-region tradeoff: at a 50~Mbit/s rate ceiling, warping has higher SSIM at 67~ms on all four scenes, whereas the proxy has higher measured SSIM than all evaluated warping configurations at 367 ms on all four scenes, and at 183 ms on three of the four scenes. %The disagreement between SSIM and FLIP on textured content further indicates that this advantage is primarily in structural completeness and viewpoint coverage rather than uniformly higher
%perceptual fidelity.

The 3DGS proxy therefore complements rather than replaces existing remote-rendering approaches. Viewpoint-specific video remains highly rate-efficient when rendering from the server pose is sufficient. When the client can execute the application, one-time asset delivery followed by state synchronization reproduces the original renderer exactly with a transmission cost that is highly dependent on the scene and assets.
%: the LZMA-compressed Unity bundle of the toy scene occupies 8.3~MB, and with \StateKbps{}~kbit/s of 60-Hz camera, object, and lighting state the amortized rate is 1.16~Mbit/s over 60~s and 0.27~Mbit/s over 300~s, below the lowest rate ceiling of Section~\ref{sec:streaming}. 
Two obstacles restrict this option: The first one is intellectual property: asset delivery places the meshes, materials, shaders, textures, and simulation logic on the client device, which content providers may not permit, whereas the proxy conveys an approximation of the observed scene without its source assets or logic. The second one is client capability: large and complex models may exceed the memory and compute budget of thin clients, such as phones and standalone XR headsets, whereas the proxy workload is governed by the streamed Gaussian representation rather than the complexity of the original renderer. The 3DGS proxy therefore targets the case in which the original renderer or assets cannot be deployed at the client but local viewpoint flexibility is valuable. This flexibility comes with additional costs: newly observed regions require time to converge, maintaining the representation can require more network traffic than native asset delivery, and the client must execute a Gaussian rasterizer.%The current relighting model captures only low-frequency diffuse transfer and approximate visibility, while dynamic geometry is restricted to engine-tracked rigid objects.

The evaluation covers four scenes and one walkthrough per scene family, with measurements performed on desktop GPUs and without wide-area network impairments.  We do not evaluate packet loss, heterogeneous or concurrent clients or substantially larger scenes. Future work should therefore evaluate the approach under realistic network conditions and at larger scale, and investigate hybrid designs in which video provides immediate high-quality coverage while a progressively converged Gaussian representation provides local six-degree-of-freedom viewpoint correction.

\section{Conclusion}
In this work, we presented a streaming system that constructs, continuously refines, and progressively streams a persistent 3D Gaussian proxy from a running game engine, enabling the client to render from its current viewpoint rather than the most recent server-rendered pose. Across four scenes, proxy quality remains stable as pose offset increases. At a 50 Mbit/s rate ceiling, the proxy has higher measured SSIM than all evaluated warping configurations at 367 ms on all four scenes, and at 183 ms on three of the four scenes. These results position persistent 3D Gaussian proxies as a view-flexible remote-rendering layer between viewpoint-specific image transport and full client-side application execution, particularly when the original renderer cannot be deployed at the client.

%\begin{acks}
%To Robert, for the bagels and explaining CMYK and color spaces.
%\end{acks}

\bibliographystyle{ACM-Reference-Format}
\bibliography{unity_3dgs_refs}

\end{document}